\documentclass[12pt]{article}
\usepackage[top=2.0cm,bottom=2.0cm,left=2.54cm,right=2.54cm]{geometry}
\usepackage{mathrsfs,graphicx,rotating,amsmath,amsfonts,mathtools,booktabs,amssymb,wasysym,caption}
\usepackage{standalone}
\usepackage{cite}
\usepackage{preview}
\usepackage{bm}
\usepackage{comment}
\usepackage{subfigure}
\usepackage{float}
\usepackage{color} 

\usepackage[hyperindex=true,
          pdfstartview=FitH,
          bookmarksnumbered=true,
          bookmarksopen=true,
          citecolor=blue,
          linkcolor=blue,
          colorlinks=true,
          pdfborder=001,
          unicode]{hyperref}

\allowdisplaybreaks
\newcommand{\rd}{{\rm d}}

\newcommand{\pfrac}[2]{\left(\frac{\partial #1}{\partial #2}\right)}
\newcommand{\bfrac}[2]{\left(\frac{#1}{#2}\right)}

\newcommand{\pfn}[3]{\left(\frac{\partial^{#3} #1}{\partial #2^{#3}}\right)}

\newcommand{\blue}[1]{\textcolor{blue}{#1}}
\renewcommand{\blue}[1]{\textcolor{black}{#1}}  
\newcommand{\red}[1]{\textcolor{red}{#1}}
\renewcommand{\red}[1]{\textcolor{black}{#1}}  
\begin{document}

\title{Restricted phase space thermodynamics 
for black holes in higher dimensions and higher curvature gravities}

\author{Xiangqing Kong\thanks{
{\em email}: \href{mailto:2120200165@mail.nankai.edu.cn}{2120200165@mail.nankai.edu.cn}},~
Tao Wang\thanks{
{\em email}: \href{mailto:taowang@mail.nankai.edu.cn}{taowang@mail.nankai.edu.cn}},~
Zeyuan Gao\thanks{{\em email}: \href{mailto:2120190129@mail.nankai.edu.cn}
{2120190129@mail.nankai.edu.cn}}~
and Liu Zhao\thanks{Correspondence author, {\em email}: 
\href{mailto:lzhao@nankai.edu.cn}{lzhao@nankai.edu.cn}}\\
School of Physics, Nankai University, Tianjin 300071, China
}

\date{}
\maketitle

\begin{abstract}
The recently proposed restricted phase space thermodynamics is shown to 
be applicable to a large class of higher dimensional 
higher curvature gravity models  coupled to Maxwell field, which are 
known as black hole scan models and are labeled by the spacetime 
dimension $d$ and the highest order $k$ 
of the Lanczos-Lovelock densities appearing in the action. 
Three typical example cases with $(d,k)=(5,1), (5,2)$ and $(6,2)$ 
are chosen as example cases and studied in some detail. These cases 
are representatives of Einstein-Hilbert, Chern-Simons and Born-Infield 
like gravity models. Our study indicates that the Einstein-Hilbert
and Born-Infield like gravity models have similar thermodynamic behaviors, e.g.
the existence of isocharge $T-S$ phase transitions with the same 
critical exponents, the existence of isovoltage $T-S$ transitions and the 
Hawking-Page like transitions, and the similar high temperature 
asymptotic behaviors for the isocharge heat capacities, etc. 
However, the Chern-Simons like $(5,2)$-model behaves quite differently. 
Neither isocharge nor isovoltage $T-S$ transitions could occur and no 
Hawking-Page like transition is allowed. This seems to indicate that the 
Einstein-Hilbert and Born-Infield like models belong to the same 
universality class while the Chern-Simons like models do not.
\end{abstract}

\section{Introduction}

Understanding the thermodynamic properties of black holes is among the central 
tasks of modern gravitational physics. Although black hole thermodynamics has 
been studied for over half a century, this field of study is far from being completed.
Different proposals/formalisms emerge from time to time, which can be briefly classified 
into two major classes, i.e. traditional black hole thermodynamics (TBHT) 
{\cite{bekenstein1972black, bekenstein1973black, bardeen1973four, hawking1975particle}} 
and extended phase space thermodynamics (EPST){\cite{kastor2009enthalpy, 
dolan2010cosmological, dolan2011pressure, dolan2011compressibility, kubizvnak2012p, 
cai2013pv, kubizvnak2017black, xu2014critical, xu2014gauss, lemos2018black, 
zhang2015phase}} for AdS black holes. In particular, Visser's recent proposal for a 
holographic thermodynamics {\cite{visser2022holographic}} (when applied to black 
holes) can be regarded as a further development of EPST, which also applies specifically 
to AdS black holes only {\cite{karch2015holographic, cong2021thermodynamics}}. 

The resent interests in black hole thermodynamics has been largely triggered by 
EPST because this formalism of black hole thermodynamics reveals some possible nontrivial 
thermodynamic behaviors of black holes such as phase transitions and critical phenomena. 
However, the EPST formalism is not free of internal issues. 
For instance, well-known Smarr relations \cite{smarr1973mass} for black hole
solutions always contain rational coefficients, which indicates
the lack of complete Euler homogeneity (see, e.g. \cite{callen1998thermodynamics} 
for standard description for the importance of Euler homogeneity in standard thermodynamics) 
and constitutes a stumbling stone in understanding 
the conditions for equilibrium conditions for black holes; the variable 
cosmological constant \blue{causes the ensemble of theories issue}; the 
interpretation of black hole mass as enthalpy looks unnatural and lacks 
physical interpretation; and the inability to extend to non-AdS black 
holes seems to indicate that the EPST formalism is in short of universality, etc. 

In some recent works, we proposed yet another formalism for black hole thermodynamics, 
i.e. the restricted phase space thermodynamics (RPST) {\cite{gao2021restricted, 
gao2022thermodynamics, wang2021black, zhao2022thermodynamics}}, which is a 
restricted version of Visser's proposal \cite{visser2022holographic} 
by removing the $P,V$ variables. Unlike the previous formalisms, 
our formalism is free of all the above mentioned issues because i) the cosmological 
constant is fixed and removed from the list of thermodynamic variables, this choice 
removes the ensemble of theories issue and brings back the interpretation of black hole 
mass from enthalpy to internal energy; ii) the introduction of the effective number of 
microscopic degrees of freedom $N=L^{D-2}/G$, together with its conjugate, the 
chemical potential $\mu= GTI_E/L^{D-2}$, where $G$ is the variable Newton 
constant, $I_E$ is the Euclidean action (which has been extensively studies 
in various cases \cite{gibbons1977action, york1986black, gibbons2005first}) and $L$ is a constant length scale, makes the 
Euler homogeneity hold perfectly; iii)  the definition of $N$ and $\mu$ as given above is
independent of holographic duality, therefore the RPST formalism is valid 
not only for AdS black holes but also for non-AdS ones. 
\red{Regarding the Euler homogeneity, let us remind that, 
there exist some works which were able to realize 
the Euler homogeneity using the EPST formalism \cite{Tian1,Tian2}, however 
those works relies heavily on the holographic duality and hence is not
applicable to non-AdS cases.}

Up till now, our study of RPST formalism is limited to black holes in Einstein gravity 
in four and higher dimensions. It remains to check whether this formalism works also 
for alternative theories of gravity, e.g. higher curvature gravities  {\cite{cai2002gauss,lanczos1932elektromagnetismus,lovelock1971einstein}}. The present work 
is a first step toward this direction. In this work, we will extend the RPST formalism 
to higher curvature gravity theories. Since the RPST formalism requires a fixed 
cosmological constant (be it positive, negative or zero), the class of 
black hole scan models \cite{crisostomo2000black} (which are a subclass of 
Lanczos-Lovelock models with some particular choices of the coupling coefficients) 
is an ideal choice for such studies, because 
such models admit a unique AdS solution with fixed cosmological constant. 
\red{Please be reminded that choosing the black hole scan models 
as the first example for the application of the RPST formalism to higher curvature gravity models does not imply that such models are 
specific. It is simply a choice for convenience because this choice 
allows us to study representatives for different types of gravity models 
starting from a single unified action.}
To illustrate the strength of this formalism, we will consider in detail the 
thermodynamic behaviors of three concrete cases, i.e. the $d=5, k=1,2$ and $d=6,k=2$ 
cases, where $d$ is the spacetime dimension and $1\leq k\leq \left[\frac{d-1}{2}\right]$. 
The reason for choosing these three particular example cases lies in that these are 
the simplest representative cases for Einstein-Hilbert (EI), Chern-Simons (CS) 
and Born-Infield (BI) gravity models without degeneracy between different classes. 
For $d=3$, the EH and CS theories degenerate, 
while for $d=4$, the EH and BI theories degenerate.
That's why we start at $d=5$. 

This paper is organized as follows. In Section 2, we briefly review the 
gravity models to be analyzed and introduce the necessary thermodynamic variables. 
It is shown that the RPST formalism applies perfectly to the whole class of 
the so-called black hole scan models \cite{crisostomo2000black}. 
The Euler relation and the first law 
hold simutaneously which implies the Gibbs-Duhem relation and Euler homogeneity. 
Section 3 is devoted to the case studies. Three concrete models will be taken as example 
cases, i.e. the $(d,k)=(5,1), (5,2)$ and $(6,2)$ models. Among these, the 
$(5,1)$-model will be studied in great detail, and various thermodynamic behaviors 
are described carefully. The $(5,2)$- and $(6,2)$-models will be treated much more 
briefly, however without sacrifice in the description of the major thermodynamic 
behaviors. In Section 4 we present the summary of the work and some further discussions.

\section{Review of the models and the RPST formalism}

Throughout this article we work in units $k_B=\hbar=c=1$, 
$\varepsilon_0 = 1/\mu_0 = 1/{A}_{d-2}$, 
where ${A}_{d-2}=\frac{2\pi^{(d-1)/2}}{\Gamma\left(\frac{d-1}{2}\right)}$ is 
the area of the unit $(d-2)$-sphere, but leave the gravitational constant intact, 
because the gravitational constant is not a constant but rather among 
the thermodynamic variables in the RPST formalism. 

The action for the black hole scan models coupled to the electromagnetic field 
can be written in two parts, 
\begin{align}
\mathcal{A}_{(d,k)}=I_{(d,k)}+I_M,\label{action}
\end{align}
where $I_{(d,k)}$ denotes the gravitational action 
and $I_M$ is the Maxwell action
\begin{align}
I_{M}=-\frac{1}{4 {A}_{d-2}} \int \sqrt{-g} F^{\mu \nu} F_{\mu \nu} \rd^{d} x.	
\end{align}

The expression for $I_{(d,k)}$ reads
\begin{align}
I_{(d,k)}=\kappa_{(d,k)} \int \sum_{p=0}^{k} c_{p}^{k} L^{(p)}, \label{Ik}
\end{align}
where $L^{(p)}$ is the $p$-th order Lanczos-Lovelock density which 
can be written in terms of the Riemann curvature two form and the vielbein $e^a$ as  
\[
L^{(p)} = \epsilon_{a_1 \cdots a_d} R^{a_1 a_2} \cdots R^{a_{2p-1}a_{2p}}
e^{a_{2p+1}}\cdots e^{a_d},
\]
where $R^{ab}=\rd\omega^{ab}+\omega^a{}_c\omega^{cd}$ and wedge product between forms 
is understood. The Lanczos-Lovelock coupling coefficients $c_{p}^{k}$ are taken to be 
\begin{align}
c_{p}^{k}= 
\begin{cases}
\dfrac{\ell^{2(p-k)}}{(d-2 p)}
\begin{pmatrix}
	k \\
	p
\end{pmatrix},
 & (p \leq k), \\
	0, & (p>k),
\end{cases}\label{coeff}
\end{align}
where the constants $\kappa_{(d,k)}$ and $\ell$ are related to the Newton  
constant $G_{(d,k)}$ and the cosmological constant via %($[G_k]=(length)^{d-2k}$)
\begin{align}
\kappa_{(d,k)}&=\frac{1}{2(d-2)!A_{d-2}G_{(d,k)}},\quad
\Lambda= -\frac{(d-1)(d-2)}{2\ell^2}. \label{lambda}
\end{align}
Notice that $c_p^k$ has dimension $[\text{length}]^{2(p-k)}$ and $G_{(d,k)}$ has
dimension $[\text{length}]^{d-2k}$, so that $c_p^k/G_{(d,k)}$ has dimension
$[\text{length}]^{2p-d}$ which is independent of $k$. For $p=1$ this fixes the 
dimension of the coupling coefficient in front of the Einstein-Hilbert action.

The above specific choices for the coupling coefficients makes 
the class of black hole scan models distinguished from the 
generic Lanczos-Lovelock models. The spacetime dimension $d$ and the highest order $k$ 
of the Lanczos-Lovelock densities appearing in the action are used to 
label the concrete model and hence the corresponding model will be referred to 
as the $(d,k)$-model. The $(d,k)$-models with $d\leq 11$ are summarized 
in \cite{crisostomo2000black}. It should be emphasized 
that each $(d,1)$-model is simply the Einstein-Hilbert theory in $d$-dimensions 
(i.e. EH model). Moreover, each $(2k+1,k)$-model is a CS model, and each  
$(2k+2,k)$-model is a BI model.

The merit for the particular choice \eqref{coeff} for the coupling coefficients lies in 
that the corresponding charged spherically symmetric AdS black hole solutions 
can be given in a unified form for 
any $d> 3$ with a fixed cosmological constant as given in eq.\eqref{lambda}. 
The metric takes the form \cite{crisostomo2000black}
\begin{align}
\rd s^{2}_{(d,k)}=-f_{(d,k)}(r) \rd t^{2}
+ \dfrac{\rd r^{2}}{f_{(d,k)}(r)}+r^{2} \rd \Omega_{d-2}^{2},
\end{align}
where $\rd \Omega_{d-2}^{2}$ is the line element on a unit $(d-2)$-sphere and
\[
f_{(d,k)}(r)=1+\frac{r^{2}}{\ell^{2}}-g_{(d,k)}(r),\quad
g_{(d,k)}(r)=\left(\frac{2 G_{(d,k)} M+\delta_{d-2 k, 1}}{r^{d-2 k-1}}
-\frac{G_{(d,k)}}{(d-3)} \frac{Q^{2}}{r^{2(d-k-2)}}\right)^{1/k},
\]
wherein $M$ and $Q$ are integration constants which are interpreted as the 
black hole mass and electric charge respectively. This metric 
is accompanied by the static electromagnetic field $A_\mu = \Phi \delta_{\mu 0}$, 
where $\Phi$ represents the Coulomb potential
\[
\Phi=\frac{1}{(d-3)} \frac{Q}{r^{d-3}}.
\]
The case $d=3$ is exceptional, because in three dimensions the Coulomb potential 
is not constant but rather logarithmic in $r$. The RPST formalism for charged AdS 
black hole solution in this exceptional case is studied in \cite{wb}.  
For $d>3$, $\Phi$ and $Q$ always have the same sign, therefore 
it suffices to consider the cases with $Q>0, \Phi>0$ in this work.

The event horizon of the black hole is located at one of the zero $r_h$ of the function
$f_{(d,k)}(r)$. Thus the mass of the black hole can be solved from 
the equation $f_{(d,k)}(r_h)=0$, yielding
\begin{align}
M(r_h,G_{(d,k)},Q)=-\frac{\delta _{1,d-2 k}}{2 G_{(d,k)}}
+\frac{r_h^{d-2 k-1} \left(\frac{r_h^2}{\ell^2}+1\right)^k}{2 G_{(d,k)}}
+\frac{Q^2}{2(d-3)r_h^{d-3}}.\label{mass}
\end{align}

The temperature of the black hole can be evaluated using the Euclidean period method, 
which gives 
\begin{align}
T&=\frac{1}{4\pi}\frac{\rd f_{(d,k)}}{\rd r}\Big|_{r=r_h}\nonumber\\
&=\frac{1}{4\pi k r_h^{2d-1}}
\left(1+\frac{r_h^2}{\ell^2}\right)^{-(k-1)} 
\left[(d-2 k-1) r_h^{2 d} \left(1+\frac{r_h^2}{\ell^2}\right)^k
-G_{(d,k)} Q^2 r_h^{2 k+4}\right]+\frac{2 r_h}{ 4 \pi \ell^2}.
\label{temp}
\end{align}
The entropy of the black hole is \cite{crisostomo2000black}
\begin{equation}\label{entropy}
S=S(r_h,G_{(d,k)},Q)=\frac{2 \pi k}{G_{(d,k)}} 
\int_{0}^{r_{h}} r^{(d-2 k-1)}\left(1+\frac{r^{2}}{\ell^{2}}\right)^{k-1} \rd r,
\end{equation}
which follows from the identity (c.f. eq.(64) in \cite{crisostomo2000black})
\begin{align}
I_E=T^{-1} M-S-T^{-1} \Phi Q, \label{IEid}
\end{align}
where $I_{E}$ is the Euclidean on-shell action. $I_E$ is presented 
in \cite{crisostomo2000black} in a complicated integration form, which will not be 
made use of in this work,
%\begin{comment}
\begin{align}
I_{E}&=-\beta \int_{r_{+}}^{\infty} \frac{N}{2}
\left[\frac{\rd}{\rd r}\left\{\frac{r^{d-1}}{G_{k}}
\left[F(r)+\frac{1}{l^{2}}\right]^{k}\right\}
-\frac{1}{\epsilon} r^{d-2} p^{2}\right] \rd r
\nonumber\\
&\quad -\frac{1}{\epsilon} \beta \int_{r_{+}}^{\infty} \Phi 
\frac{\rd}{\rd r}\left(r^{d-2} p\right) \rd r+B_{E}.
\end{align}
%\end{comment}
For our purpose, it is important {\em not} to use the above identity for 
defining the entropy. Rather, the entropy should be calculated using 
Bekenstein-Hawking formula (when applicable) or 
Wald method {\cite{wald1993black,iyer1994properties}}. It turns out that 
the same result \eqref{entropy} arises, at least 
in the concrete cases to be described in detail in the next section.

In order to establish the RPST for the $(d,k)$-models, we need to introduce a novel pair of
thermodynamic variables, i.e. the effective number of microscopic degrees of freedom $N$ and
the corresponding chemical potential $\mu$. These two quantities are defined as
\begin{align}
\blue{N=N(G_{(d,k)})=\frac{L^{d-2k}}{G_{(d,k)}}},\quad 
\mu=\mu(r_h,G_{(d,k)},Q)=\frac{T I_E}{N}, 
\end{align}
where $L$ is a constant length scale which is introduced to make 
$N$ dimensionless. It should be noticed that $G_{(d,k)}$ is {\it not} the standard 
Newton constant $G$ but rather differs from $G$ by a multiplicative constant factor. 
This slight difference does not make any harm to the definitions of $N,\mu$, because 
the multiplicative constant factor can be absorbed by  a simple redefinition of the 
arbitrary constant length scale $L$. However, when evaluating the black hole entropy
using Bekenstein-Hawking formula {\cite{hawking1975particle}} (when applicable), the multiplicative constant factor
becomes relevant and needs to be taken care of.   

Another necessary step before establishing 
the RPST formalism is to make a rescaling for the electric charge and the Coloumb 
potential, so that $\kappa_{(d,k)}$ becomes an overall factor in front of the total action. 
The rescaled charge and potential read
\begin{align}
\hat{Q}&=\hat{Q}(Q,G_{(d,k)})=\frac{Q L^{(d-2k)/2}}{\sqrt{G_{(d,k)}}},
\label{qhat}\\
\hat{\Phi}&=\hat{\Phi}(r_h,G_{(d,k)},Q)=\frac{\Phi \sqrt{G_{(d,k)}}}{L^{(d-2k)/2}}
=\frac{ \sqrt{G_{(d,k)}} Q }{(d-3)L^{(d-2k)/2} r_h^{d-3}}.
\label{Phihat}
\end{align}

After the above preparation, it is now straightforward to recognize that eq.\eqref{IEid}
is nothing but the Euler relation
\begin{align}
M=TS+\hat{\Phi}\hat{Q}+\mu N, \label{euler}
\end{align}
and consequently it is easy to check that total differential of the black hole mass 
obeys the following first law of 
thermodynamics,
\begin{align}
\rd M=T\rd S+\hat{\Phi}\rd \hat{Q}+\mu \rd N. \label{1stlaw}
\end{align}
Eqs.\eqref{euler} and \eqref{1stlaw} indicate that the RPST formalism is applicable to 
$(d,k)$-models for any $d> 3$ 
and any admissible $k$. It is worth mentioning that the Euler relation 
\eqref{euler} and the first law \eqref{1stlaw} take exactly the same form for any admissible 
$(d,k)$-models. This is in contrast to the EPST formalism in which the Euler relation is 
always absent and the number of thermodynamic variables can be different in different 
models (e.g. the inclusion of coupling coefficients in the set of thermodynamic variables).
Moreover, eqs. \eqref{euler} and \eqref{1stlaw} imply that $M$ is a first order homogeneous
function in the extensive variables $S, \hat Q, N$ while the intensive variables 
$T,\hat\Phi,\mu$ are zeroth order homogeneous functions in the extensive variables. Combining 
eqs. \eqref{euler} and \eqref{1stlaw} we can also obtain the Gibbs-Duhem relation
\[
S\rd T+\hat Q\rd\hat\Phi+N\rd\mu=0,
\]
which is an important relationship between the intensive variables in order that 
the thermodynamic system is consistent.

\section{Case studies and thermodynamic behaviors}

Although the form of the Euler relation and the first law remains the same for any 
$(d,k)$-models, the thermodynamic behaviors can be drastically different for
different choices of $(d,k)$. In this section, three concrete cases will be studied 
in detail, each corresponds to a different type of gravity models. The chosen cases 
are, respectively, the $(5,1)$-model (EH gravity), the $(5,2)$-model (CS gravity) 
and the $(6,2)$-model (BI gravity). Each model will be described in a separate subsection.

\subsection{$(5,1)$-model: EH gravity coupled to Maxwell field}

\subsubsection{Description of phase transitions}

For all $(d,1)$-models the thermal and caloric equations of states for the black holes 
can be worked out analytically by use of eqs.\eqref{mass}-\eqref{Phihat}. 
This can be achieved by solving $r_h, G_{(d,k)}, Q$ from the expressions 
$S=S(r_h,G_{(d,k)},Q), N=N(G_{(d,k)}), \hat{Q}=\hat{Q}(Q,G_{(d,k)})$ 
which yields
\begin{align}
\blue{r_h=(2 \pi )^{-\frac{1}{d-2}} 
\left(\frac{(d-2)  L^{d-2} S}{N}\right)^{\frac{1}{d-2}},\quad
G_{(d,1)}=\frac{L^{d-2}}{N},\quad 
Q=\frac{\hat Q}{\sqrt{N}}.}
\label{rhGQ}
\end{align}
Substituting the above solution into eqs.\eqref{mass} and \eqref{temp}, 
we get
\begin{align}
T(S,\hat Q,N)&=\frac{1}{2}(2\pi)^{\frac{3-d}{d-2}} \tilde{S} 
\left(d-3+\frac{(d-1) (2 \pi )^{-\frac{2}{d-2}} }{\ell^2} \tilde{S}^2
-(2 \pi )^{\frac{2(3-d)}{2-d}}L^{d-2}\frac{\tilde{S}^{2(3-d)}\hat{Q}^2}{N^2}\right),
\label{thermaleos}\\
M(S,\hat Q,N)&=\frac{1}{2}(2 \pi )^{\frac{d-3}{2-d}}  L^{2-d} N \tilde{S}^{d-3} 
\left(1+\frac{(2 \pi )^{-\frac{2}{d-2}}}{\ell^2}\tilde{S}^2
+\frac{(2 \pi )^{\frac{2(3-d)}{2-d}}L^{d-2}}{d-3}
\frac{\tilde{S}^{2(3-d)} \hat{Q}^2}{N^2}\right),
\end{align}
where 
\begin{align}
\tilde{S}=(d-2)^{\frac{1}{d-2}} L\left(\frac{S}{N}\right)^{\frac{1}{d-2}}.
\end{align}
It is transparent that $M$ behaves as $M\to \lambda M$ when the extensive variables
$S, \hat Q, N$ are rescaled as $S\to \lambda S, \hat Q\to\lambda \hat Q, N\to \lambda N$,
while $T$ is kept unchanged under the above rescaling. The zeroth order homogeneity of $T$
is very important in understanding the condition for thermal equilibrium.

It can also be shown that for any $d>3$, the thermal equation of states \eqref{thermaleos}
always contains an inflection point with 
\begin{align}
\pfrac{T}{S}_{\hat Q}
=\pfrac{^2 T}{S^2}_{\hat Q}=0, \label{inflect}
\end{align}
which indicates that the $T-S$ phase structure is 
similar for all EH models in $d>3$. Therefore, it suffices to consider
only the case $d=5$ as a representative for all EH models with $d\geq 4$.

To be more explicit, let us rewrite the full action for the $(5,1)$-model as follows,
\begin{align}
\mathcal{A}_{(d,k)}
%\kappa_1 \int (c_0^1 L^{(0)} + c_1^1 L^{(1)})+I_M
%=\frac{1}{2(d-2)!A_{d-2}G_{(5,1)}} \int (\frac{l^{-2}}{d} d!+\frac{l^0}{d-2} (d-2)!R)+I_M\\
&=\frac{1}{12 \pi^{2} G_{(5,1)}}  \int\left(R+\frac{12}{\ell^{2}}\right) \sqrt{-g} \rd^5 x 
-\frac{1}{8\pi^2} \int F_{\mu\nu}F^{\mu\nu} \sqrt{-g} \rd^5 x.
\end{align}
This is nothing but the Einstein-Maxwell theory in five dimensions, with the corresponding
charged spherically symmetric black hole solution given by the metric
\begin{align}
\rd s^{2}_{(5,1)}=-f(r) \rd t^{2}+\frac{\rd r^{2}}{f(r)}+r^{2} \rd \Omega_{3},
\end{align}
wherein
\[
f(r)=\left(1+\frac{r^{2}}{\ell^{2}}
+\frac{G_{(5,1)} Q^{2}}{\pi r^{4}}-\frac{2G_{(5,1)} M}{r^{2}}\right).
\]
Since this is Einstein-Hilbert gravity, the entropy of the black hole can be evaluated 
by simply employing the Bekenstein-Hawking formula $S=\frac{A}{4G}$, where $G$ is the 
Newton constant which appear in the standard Einstein-Hilbert action 
$\frac{1}{16\pi G}\int R\sqrt{-g}\rd^d x$. It is evident that 
\[
\frac{1}{16\pi G}=\frac{1}{12 \pi^{2} G_{(5,1)}},\quad{\rm i.e.}\quad 
G =\frac{3\pi G_{(5,1)}}{4},
\]
thus
\[
S=\frac{A_3 r_h^3}{4G} = \frac{2\pi r_h^3}{3G_{(5,1)}}.
\]
This result coincide with what can be obtained by evaluating the integral 
in eq.\eqref{entropy}.

By solving the inflection point equation \eqref{inflect} in $d=5$, we get the 
critical point parameters
\begin{align}
S_c=\frac{2 \pi\ell^3}{9 \sqrt{3} L^3 } N,\quad
\hat{Q}_c=\frac{1}{3} \sqrt{\frac{2 }{15}}\frac{\ell^2}{L^{3/2}} N.
\end{align}
Inserting these parameters back into eqs.\eqref{thermaleos} and \eqref{Phihat}, we get
\begin{align}
T_c=\frac{4 \sqrt{3}}{5 \pi  \ell}, \quad 
\hat{\Phi}_c=\frac{1}{\sqrt{30} L^{3/2}}.
\end{align}
These critical parameters obey the identy
\[
\frac{S_c T_c}{\hat{\Phi}_c \hat{Q}_c}=8.
\]
Moreover, if we introduce the Helmholtz free-energy
\[
F=F(T,\hat Q,N)=M-TS,
\]
then the critical value of the black hole mass and Helmholtz free energy read
\begin{align}
M_c= \frac{7\ell^2}{30L^3}N,
\quad
F_c=\frac{\ell^2}{18 L^3} N.
\end{align}

Now we are in a position to study the thermodynamic behavior of the $(5,1)$-model. 
For this purpose, it will be advantageous to adopt the relative state parameters/functions
\begin{align}
s\equiv \frac{S}{S_c},\quad t\equiv \frac{T}{T_c},\quad
q\equiv \frac{Q}{Q_c},\quad \phi=\frac{\hat\Phi}{\hat\Phi_c},\quad
m\equiv \frac{M}{M_c},\quad f\equiv \frac{F}{F_c}
\label{sparam}
\end{align}
which are independent of the scale of the black hole. Then the thermal, caloric 
equations of states and the Helmholtz free energy can be written as
\begin{align}
t&= \frac{10 s^2+15 s^{4/3}-q^2}{24 s^{5/3}},\label{equationofstate}\\
m&=\frac{5 \left(s^2+3 s^{4/3}\right)+q^2}{21 s^{2/3}},\label{ceos}\\
f&= \frac{-s^2+3 s^{4/3}+q^2}{3 s^{2/3}}, \label{freef}
\end{align}
where the dependence of $f$ upon $t$ is implicit via eq.\eqref{equationofstate}. 
Notice that only two relative parameters $s, q$ appear on the right hand side of 
the above equations. 
The scale independence of the above equations is known as the law of corresponding states 
in traditional thermodynamics for ordinary matter, however in black hole thermodynamics, 
this law is only known to exist in the RPST formalism. The $\hat\Phi-\hat Q$ 
equation of states can also be re-expressed in terms of the relative parameters 
given above. The result reads
\begin{align}
\phi= \frac{q}{s^{3/2}}.	\label{phiqeq}
\end{align}

As an immediate consequence of 
the law of corresponding states, the effective number $N$ of microscopic degrees of freedom 
of the black hole is irrelevant while analyzing the thermodynamic processes. As a bonus, 
we can also see that $\ell$ and $L$ also disappear in 
eqs.\eqref{equationofstate}-\eqref{phiqeq}, 
indicating that the choice of values for these two constant parameters
is also irrelevant to the analysis of thermodynamic behaviors.

\begin{figure}[ht]
\begin{center}
\includegraphics[width=.33\textwidth]{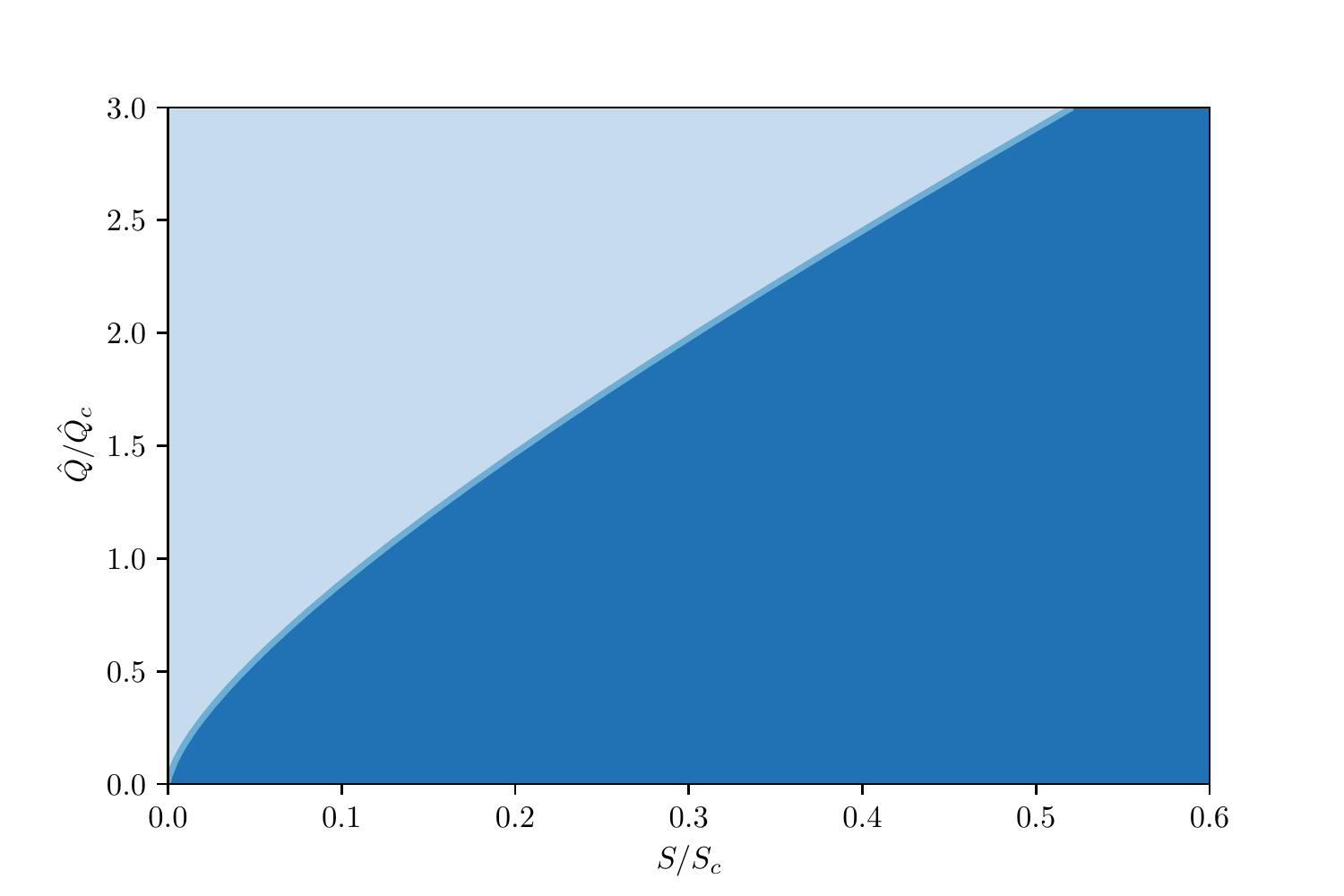}%
\includegraphics[width=.33\textwidth]{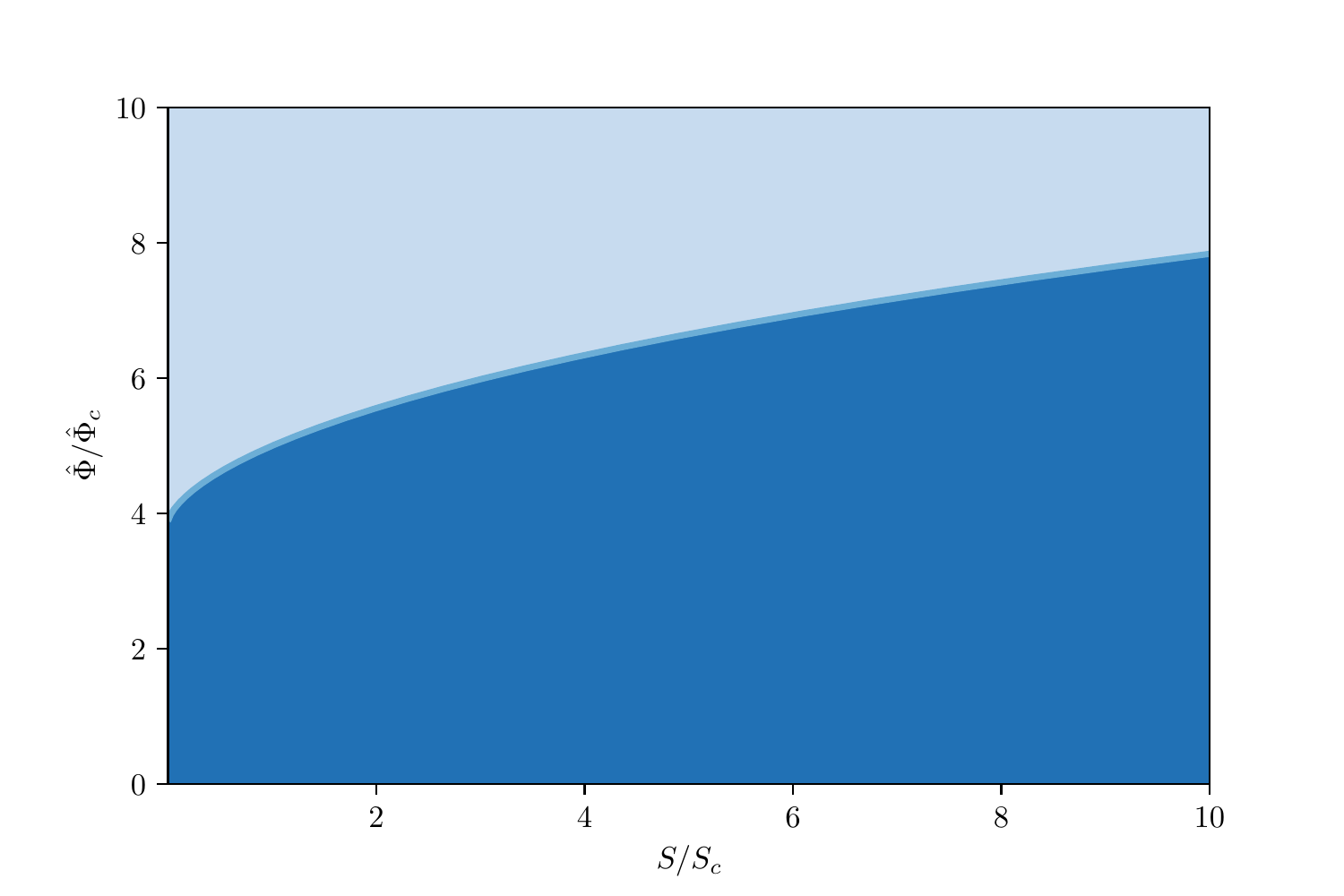}%
\includegraphics[width=.33\textwidth]{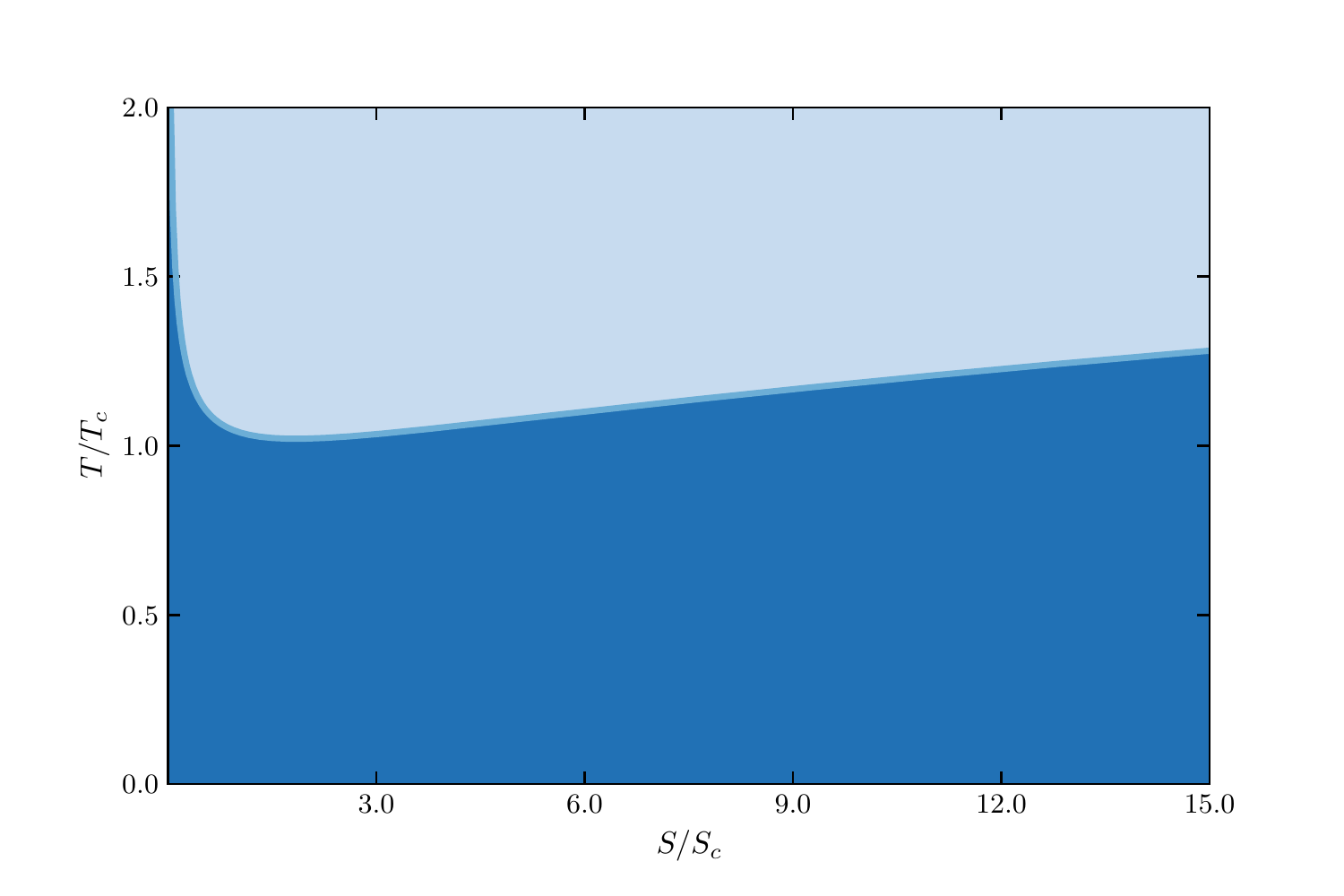}
\caption{Bounds in the parameter space}\label{bounds}
\end{center}
\end{figure}

Before analyzing the thermodynamic processes, it is important to realize that there are some 
natural bounds in the parameter space which is triggered by the requirement 
that $t\geq 0$ and that $q$ must be real-valued. It follows from eq.\eqref{equationofstate}
that $t\geq 0$ requires that
\begin{align}
q\leq \sqrt{5}\sqrt{3s^{4/3}+2s^2}, \label{qsbounds}
\end{align}
or, after employing eq.\eqref{phiqeq}, that
\[
\phi\leq \sqrt{5} \sqrt{2 s^{2/3}+3}.
\]
These bounds are nothing but the well-known Bogomol'nyi bound {\cite{bogomol1976stability}} rewritten in terms of 
the relative thermodynamic variables. In addition, from 
eq.\eqref{equationofstate}, one can solve $q$ in terms of $s,t$,
which reads 
\[
q=\sqrt{-24 s^{5/3} t+15 s^{4/3}+10 s^2}.
\] 
Thus the real-valuedness of
$q$ poses a bound for $t$ versus $s$,
\[
t\leq \frac{5 \left(2 s^{2/3}+3\right)}{24 \sqrt[3]{s}}.
\]
The allowed parameter space specified by the above bounds is shown as the dark shaded
area in Fig.\ref{bounds}.

Now let us proceed to analyze some of the thermodynamic processes.

\begin{figure}[ht]
\begin{center}
\includegraphics[width=.4\textwidth]{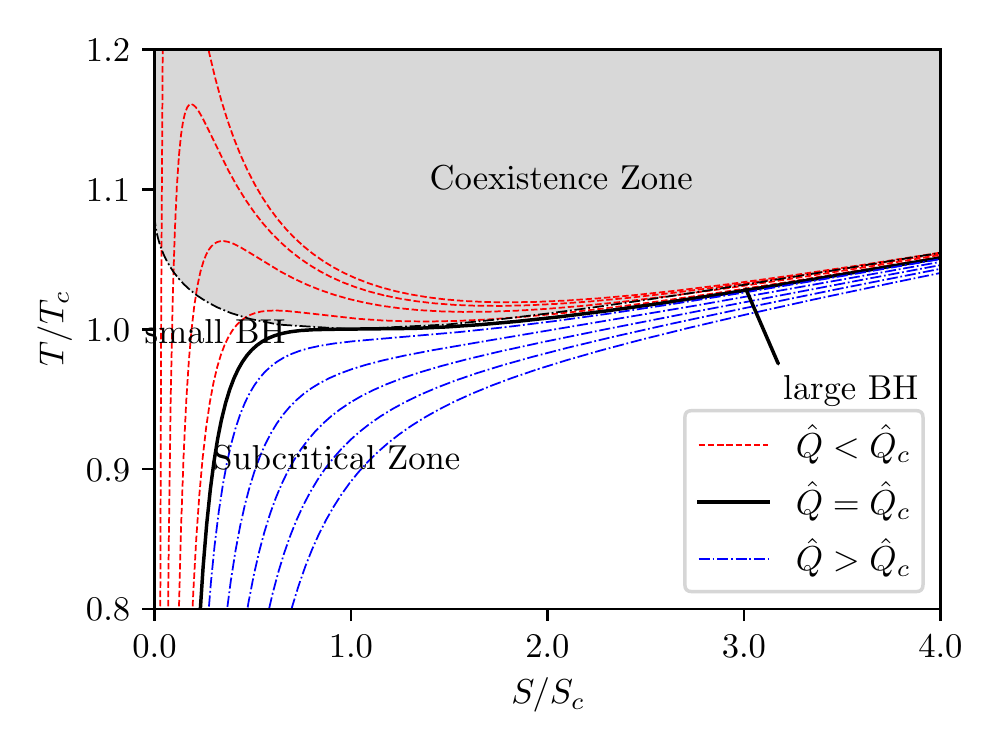}\hspace{2em}
\includegraphics[width=.4\textwidth]{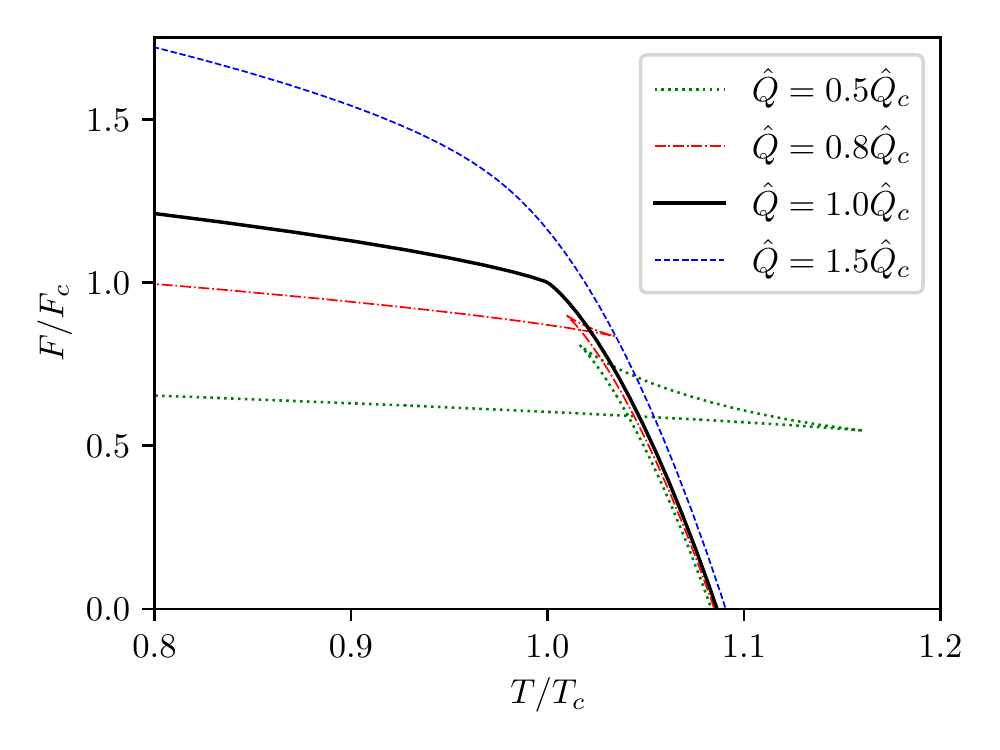}
\caption{Isocharge $T-S$ and $F-T$ curves}
\label{fig2}
\end{center}
\end{figure}

First we consider the isocharge $T-S$ processes. 
The isocharge $T-S$ and $F-T$ curves are depicted in Fig.\ref{fig2}
by use of eqs.\eqref{equationofstate} and \eqref{freef}. It can be seen that, 
at $\hat Q< \hat Q_c$, every isocharge $T-S$ curve contains a segment on 
which $T$ decreases as $S$ increases. This non-monotonic behavior implies a 
first order phase transition which occurs at a constant temperature 
determined by requiring minimization of the Helmholtz free energy, i.e.
\[
\pfrac{F}{T}_{\hat Q}=0.
\]
Effectively, this condition requires that the actual isocharge process does not 
follow the analytical curve for $\hat Q< \hat Q_c$ but rather undergoes an isocharge
isothermal segment during the phase transition at an invariant temperature $T^\ast>T_c$. 
Such phase transitions were also observed previously in our works {\cite{gao2021restricted,gao2022thermodynamics}} and are
referred to as supercritical phase transitions which occur only in black hole 
thermodynamics but not in thermodynamics of ordinary matter. During this isocharge
isothermal stage of evolution, the Helmholtz free energy remains at a fixed value
which corresponds to the root of the swallow tail depicted in the right plots in 
Fig.\ref{fig2}. The isocharge isothermal segment on the $T-S$ curve corresponds to 
the coexistence phase between the small and large black hole phases. The coexistence 
zone on the $T-S$ phase plane is shown by the shaded area on the left picture
of Fig.\ref{fig2}. At $T=T_c$, the $T-S$ phase transition becomes second order 
which corresponds to the critical point.

We can also consider the isovoltage $T-S$ processes instead of the isocharge processes. 
To do so, we need to carry out the following steps. First we re-express 
$\hat Q$ in terms of $\hat \Phi$ and $r_h$ using 
eq.\eqref{Phihat} and then using the expression for $r_h$ given in
eq.\eqref{rhGQ} to rewrite $\hat Q$ as a function in $\hat \Phi, S, N$, i.e.
\[
\hat Q= \frac{18^{1/3} L^2 \hat{\Phi} S^{2/3}}{\pi ^{2/3} {N}^{1/6}}.
\]
Inserting the above result into eq.\eqref{thermaleos} with $d=5$, we get
\[
T(S,\hat{\Phi},N)=\frac{\ell^2 ({3\pi^2 N^2} )^{1/3}   
\left(1-2 L^3 \hat{\Phi}^2\right)+3 L^2 (2S^2)^{1/3}}{\ell^2 L(36\pi^4 N S)^{1/3}}.
\]
We wish to analyze the $T-S$ relation at fixed $\hat\Phi$. For this purpose we 
first need to find out the extrema of the $T-S$ relation at constant $\hat\Phi, N$.
It turns out that, for $\hat\Phi$ below the threshold value 
\begin{align}
\hat\Phi_{\rm thr}
=\bfrac{1}{2L^3}^{1/2}=\sqrt{15}\hat\Phi_c >0,
\label{phithr}
\end{align}
there is a single minimum which can be determined by
$\displaystyle \pfrac{T}{S}_{\Phi,N}=0$, with the coordinate at the minimum given by
\begin{align}
S_{\rm min} = \frac{\pi \ell^3 N \left(1-2 L^3 \hat{\Phi}^2\right)^{3/2}}{3 \sqrt{2} L^3},
\quad
T_{\rm min}=\frac{\sqrt{2(1-2 L^3 \hat{\Phi}^2)}}{\pi  \ell}.
\label{STmin}
\end{align}
Introducing the relative parameters $\tilde s=S/S_{\rm min}, \tilde t={T}/{T_{\rm min}}$, 
the isovoltage $T-S$ relation can be recast in the form
\[
\tilde t =\frac{{\tilde s}^{2/3}+1}{2 \sqrt[3]{\tilde s}}.
\]
It is remarkable that the rescaled isovoltage $T-S$ relation does not depend on 
the potential $\hat\Phi$ at all provided $\hat\Phi<\hat\Phi_{\rm thr}$. 

For $\hat\Phi\geq\hat\Phi_{\rm thr}$, however, the minimum 
on the isovoltage $T-S$ curve disappears and $T$ is monotonically increasing with $S$, 
which starts at some nonvanishing zero point value which corresponds to $T=0$. 
The nonvanishing zero point value of $S$ corresponds to an extremal black hole remnant.

Similar considerations can also be applied in considering the isovoltage $\mu-T$ relation,
where the expression for $\mu$ can be inferred from the Euler relation \eqref{euler}. 
It turns out that the isovoltage $\mu-T$ relation is branched
for $\hat\Phi<\hat\Phi_{\rm thr}$, and the chemical 
potential $\mu_{\rm lower}$ in the lower branch becomes zero at the temperature
\[
T_0=\frac{3\sqrt{1-2 L^3 \hat{\Phi}^2}}{2\pi\ell},
\]
i.e. $\mu_{\rm lower}(T_0)=0$. 
Meanwhile, at the temperature $T_0$, the chemical potential $\mu_{\rm upper}$ in 
the upper branch takes the value 
\[
\mu_{\rm upper}(T_0)=\frac{\ell^2 \left(1-2 L^3 \hat{\Phi}^2\right)^2}{32 L^3}.
\]
Therefore, we can take the liberty to define the rescaled temperature $\tau$ and 
chemical potential $m$ as
\[
\tau=\frac{T}{T_0},\quad
\nu=\frac{\mu}{\mu_{\rm upper}(T_0)}.
\]
Using these rescaled parameters, the isovoltage $\mu-T$ relations can be cast in 
the following simple form,
\begin{align}
\nu&=\pm \frac{1}{8}\left(\sqrt{9 \tau^{2}-8}-3 \tau\right)^{2}
\left[\sqrt{9 \tau^{2}-8} \tau\mp (3 \tau^{2}-4)\right],
\end{align}
where the upper/lower signs correspond to the upper/lower branch of the $\mu-T$ relation.

For $\hat\Phi\geq\hat\Phi_{\rm thr}$, the branched behavior for $\mu$ disappears and
$\mu$ becomes identically negative and is monotonically decreasing with $T$.
Such behavior is similar to ordinary ideal gas,  
which looks less interesting than the case $\hat\Phi<\hat\Phi_{\rm thr}$.

\begin{figure}[ht]
\begin{center}
\includegraphics[width=.4\textwidth]{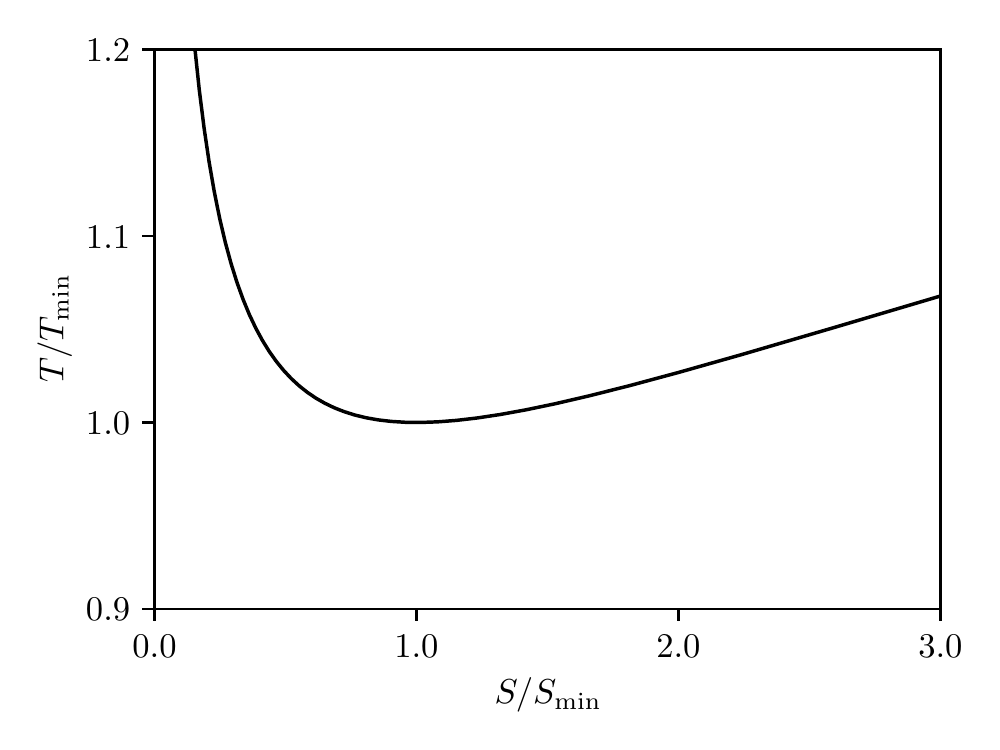}\hspace{2em}
\includegraphics[width=.4\textwidth]{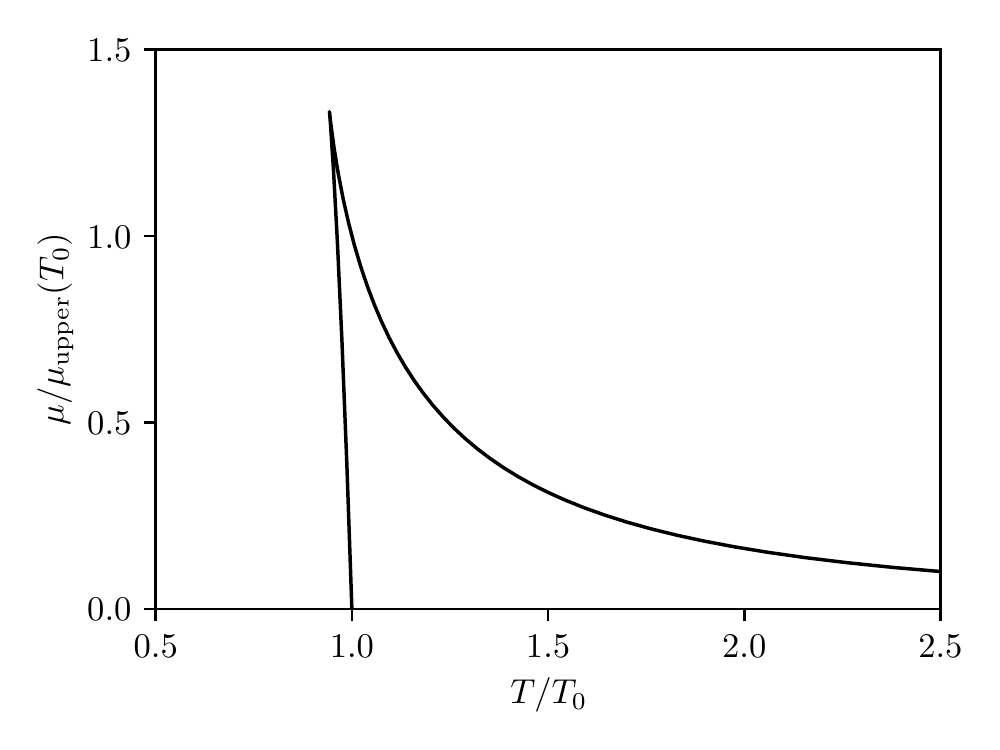}
\caption{Isovoltage $T-S$  and $\mu-T$ curves}\label{fig3}
\end{center}
\end{figure}

The isovoltage $T-S$ and $\mu-T$ curves at $\hat\Phi<\hat\Phi_{\rm thr}$ are 
shown in Fig.\ref{fig3}. 
It can be seen from the plots that for each $T>T_{\rm min}$, there are two 
black hole states, and the one with bigger entropy is thermodynamically preferred,
because only for the large black hole state with bigger entropy the temperature 
and the entropy is positively correlated, implying a positive heat capacity. 
Unlike the isocharge processes described earlier, there is no phase
equilibrium condition for the transition from the unstable
small black hole to the stable large black hole. Another point to be noticed is that, 
in the absence of Coulomb potential, the zero $T_0$ for the chemical potential 
corresponds to the famous Hawking-Page phase transition {\cite{hawking1983thermodynamics}}, which is a phase transition
from an AdS black hole to a thermal gas. The Hawking-Page temperature is found to be
\[
T_{\rm HP}= T_0|_{\hat\Phi=0} = \frac{3}{2\pi\ell}.
\]
One can check that the radius of the black hole event horizon at the 
Hawking-Page temperature is precisely the AdS radius, i.e. $r_h (T_{\rm HP})=\ell$.
In the presence of nontrivial Coulomb potential, the zero for the chemical potential
still appears, however it is not clear what physical state does the zero 
$\mu_{\rm lower}(T_0)=0$ correspond.

\begin{figure}[ht]
\begin{center}
\includegraphics[width=.4\textwidth]{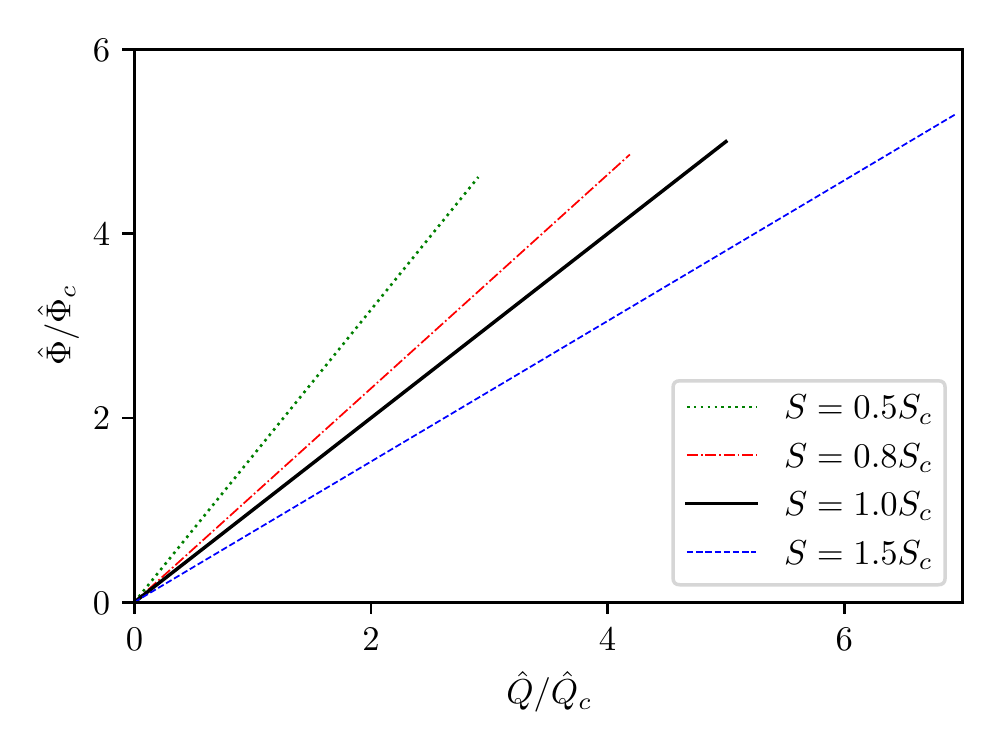}\hspace{2em}
\includegraphics[width=.4\textwidth]{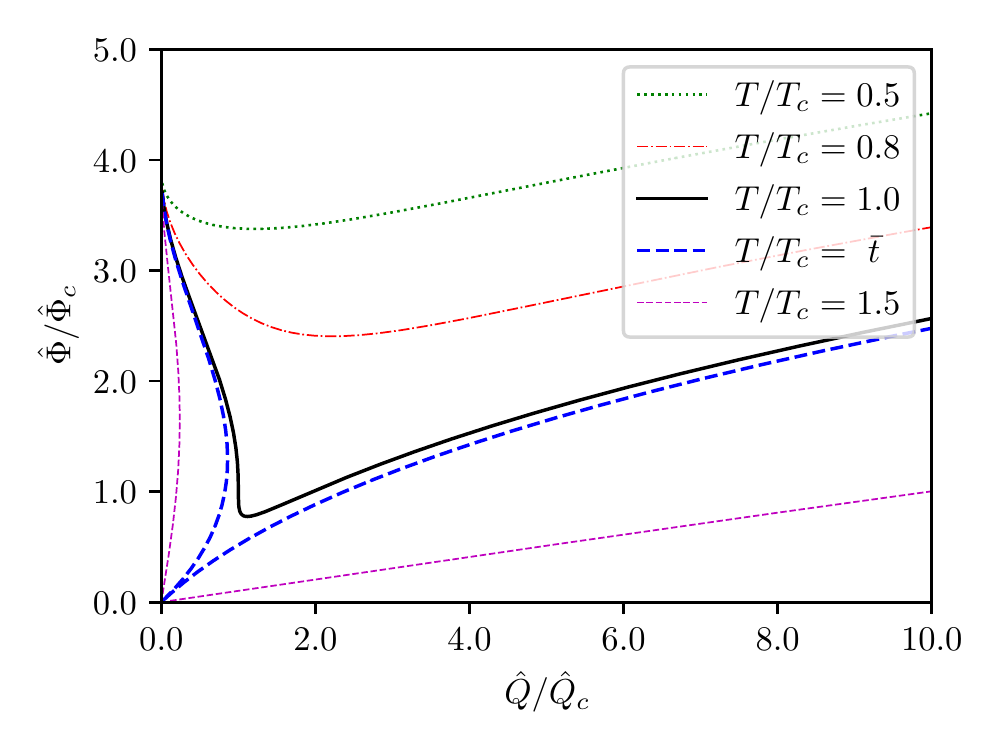}
\caption{Adiabatic and isothermal $\hat{\Phi}-\hat{Q}$ curves}\label{phi-q}
\end{center}
\end{figure}

Next let us consider the $\hat\Phi-\hat Q$ processes. There are two possible conditions
for such processes, i.e. adiabatic and isothermal, with the corresponding 
curves presented in Fig.\ref{phi-q}. The adiabatic $\hat\Phi-\hat Q$ 
processes are quite simple. It can be seen from eq.\eqref{phiqeq} that in the adiabatic 
processes, $\hat\Phi$ is always proportional to $\hat Q$. The only thing to be 
noticed is that at any constant $S$, there is an upper bound for $\hat Q$ according to 
eq.\eqref{qsbounds}. The situation for the isothermal $\hat\Phi-\hat Q$ processes is, 
however, more complicated. One needs to consider $t$ as an implicit parameter 
in the equation of states \eqref{phiqeq} and consequently the isothermal
$\hat\Phi-\hat Q$ curves are not monotonic and even not 
always single-valued. At sufficiently low temperatures 
the isothermal $\hat\Phi-\hat Q$ relation is single-valued and possesses a single minimum
at nonzero $\hat Q$, which is given by the condition $\pfrac{\hat\Phi}{\hat Q}_T=0$. 
In terms of the relative parameters, the minimum is located at
\begin{align}
q_{\rm min}= \frac{36\sqrt{15}t^2\sqrt{25-24t^2}}{125},\quad
\phi_{\rm min}= \dfrac{\sqrt{15}}{5}\sqrt{25-24\, t^{2}}.
\end{align}
These can be easily translated into the original $\hat Q,\hat\Phi$ values by 
multiplication with $Q_c$ and $\hat\Phi_c$, i.e.
\begin{align}
\hat Q_{\rm min}= q_{\rm min} Q_c,\quad
\hat\Phi_{\rm min} =\phi_{\rm min} \hat\Phi_c.
\end{align}
When $t=1$, the isovoltage $\hat\Phi-\hat Q$ curve has a single vertical tangent line. 
For $1<t< \bar t\equiv \bfrac{25}{24}^{1/2}$, the $\hat\Phi-\hat Q$ relation 
becomes multivalued and each of the corresponding curves possesses two vertical 
tangent lines. If $t\geq \bar t$, all the $\hat\Phi-\hat Q$ curves are still multivalued 
and all such curved will intersect at $(\hat Q, \hat\Phi)=(0,0)$. 
Another interesting feature of the isovoltage $\hat\Phi-\hat Q$ curves 
which is worth to be noticed is that, for any $T>0$, the corresponding 
curves always intersect at $(\hat Q, \hat\Phi)=(0,\hat\Phi_{\rm thr})$, 
where $\hat\Phi_{\rm thr}$ is defined in eq.\eqref{phithr}. The reason 
for the appearance of this intersection point remains unclear to us.

\begin{figure}[ht]
\begin{center}
\includegraphics[width=.4\textwidth]{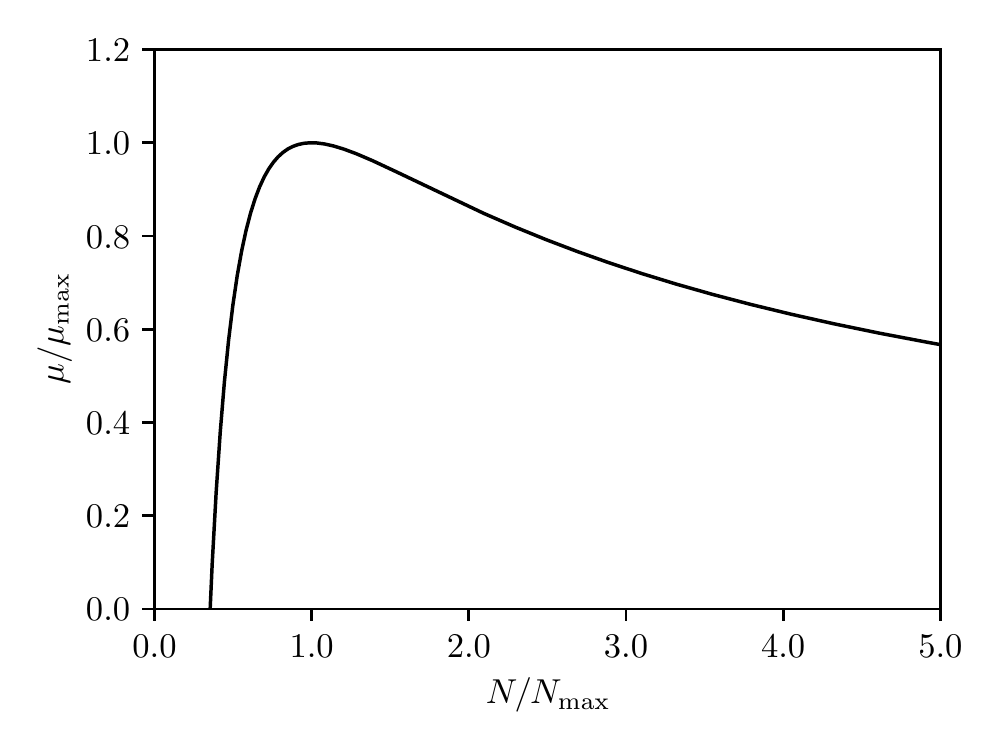}
\caption{$\mu-N$ curve} \label{fig4}
\end{center}
\end{figure}

It remains to consider the $\mu-N$ relationship in this concrete model. By use of 
the Euler relation \eqref{euler}, the chemical potential can be written as
\[
\mu=\frac{\ell^2 \left[3  (12 \pi^2 )^{1/3} L N^{2/3} S^{4/3}-2 \pi ^2 \hat{Q}^2\right]
-9 L^3 S^2}{12 ({18\pi^4})^{1/3}   \ell^2 L^2 S^{2/3} N^{4/3}}.
\] 
It can be shown that the $\mu-N$ curve
with fixed $S,~\hat Q$ does not contain any inflection point but rather has a maximum,
with the parameters at the maximum given by 
\begin{align}
N_{\rm max}=\frac{\sqrt{2} \left(2 \pi ^2 \ell^2 
\hat{Q}^2+9 L^3 S^2\right)^{3/2}}{9 \pi  \ell^3 L^{3/2} S^2},\quad
\mu_{\rm max}=\frac{3 \ell^2 S^2}{16 \pi^2 \ell^2 \hat{Q}^2+ 72L^3 S^2}.
\end{align}
Therefore, in terms of the relative parameters $u=\mu/\mu_{\rm max},~n=N/N_{\rm max}$, the 
$\mu-N$ equations of states can be written as
\begin{align}
u=\frac{2 n^{2/3}-1}{n^{4/3}}.
\end{align}
Once again, only a single relative parameter $n$ appears on the right hand side,
which can be understood as the law of corresponding states 
at an enhanced level. The corresponding $\mu-N$ curve is depicted in Fig.\ref{fig4}.

Summarizing this subsection, we have found that the thermodynamic behavior for the
charged AdS black hole solution of the $(5,1)$-model in the RPST formalism 
is quantitatively slightly 
different from that of the four dimensional RN-AdS black hole, however the qualitative 
behaviors are basically identical. The RPST behaviors for the black holes are 
similar to that of the ordinary matter in some respects, e.g. the form of the 
first law and the Euler homogeneity behaviors are similar, however there are also 
some significant differences, e.g. the appearance of supercritical equilibrium 
phase transitions in the isocharge processes, the non-equilibrium phase transitions in the
isovoltage processes and the existence of the yet-to-be-understood charged states 
with vanishing chemical potential, etc. These behaviors seems to be universal 
for different black hole solutions in Einstein gravity.

\subsubsection{Process parameters}

In the study of thermodynamic behaviors of ordinary matter, the parameters 
that characterize certain physical processes are important objects. 
The relevant parameters include the isochoric heat capacity $C_V$, the 
isobaric expansion coefficient $\alpha_P$, 
the isochoric pressure coefficient $\beta_V$ and the isothermal compression 
coefficient $\kappa_T$, etc.

In the case of charged black holes, we can introduce analogous concepts and study their
behaviors. Remember that, instead of $P-V-T$ systems, the charged black holes are
$\hat\Phi-\hat Q-T$ systems. Therefore, the analogy to isochoric heat capacity 
$C_V$ is now changed into
\[
C_{\hat Q}\equiv\pfrac{M}{T}_{\hat Q}=T\pfrac{S}{T}_{\hat{Q}},
\]
while the analogies to $\alpha_P$, $\beta_V$ and $\kappa_T$ are given, respectively, as
\begin{align}
\alpha_{\hat{\Phi}} \equiv \frac{1}{{\hat{Q}}}\pfrac{\hat{Q}}{T}_{\hat{\Phi}},\quad
\beta_{\hat{Q}} \equiv \frac{1}{\hat{\Phi}}\pfrac{\hat{\Phi}}{T}_{\hat{Q}},\quad
\kappa_T\equiv -\frac{1}{{\hat{Q}}} \pfrac{\hat{Q}}{\hat{\Phi}}_T.
\label{abc}
\end{align}
Clearly, these parameters obey the following simple identity,
\[
\alpha_{\hat{\Phi}} =\hat{\Phi} \beta_{\hat{Q}} \kappa_T.
\]

\begin{figure}[ht]
\begin{center}
\includegraphics[width=.4\textwidth]{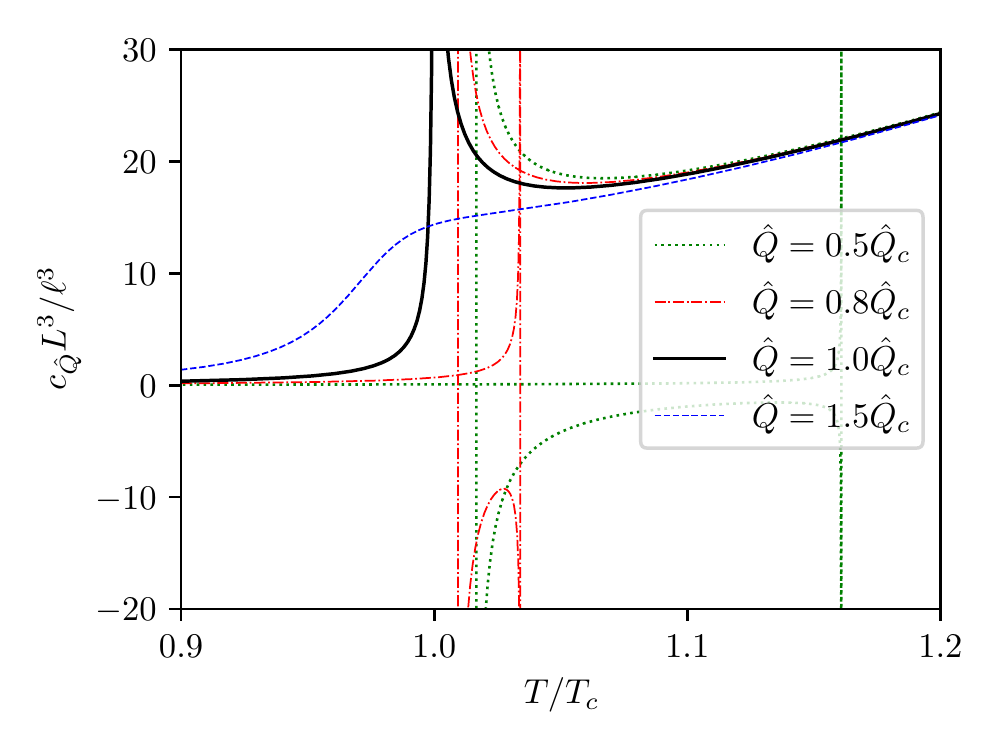}
\caption{Isocharge $c_{\hat{Q}}-T$ curves}\label{fig5}
\end{center}
\end{figure}

\noindent We also introduce the isocharge specific heat capacity $c_{\hat Q}$ as 
\[
c_{\hat Q}=C_{\hat Q}/N,
\]
because this object is scale independent and can be expressed in terms of the parameters 
given in eq.\eqref{sparam} as
\[
c_{\hat Q}=\frac{2 \pi \ell^3 s \left(-q^2+15 s^{4/3}+10 s^2\right)}{15 \sqrt{3} L^3  \left(q^2-3 s^{4/3}+2 s^2\right)}.
\]
The dependence on $s$ is explicit, while dependence on $t$ is implicit by use of eq.
\eqref{equationofstate}. 

The $c_{\hat Q}-T$ behavior is shown in Fig.\ref{fig5}. It can be seen that the isocharge 
$c_{\hat{Q}}-T$ curves for $\hat Q<\hat Q_c$ always have two divergent points, 
each corresponds to one of the two extrema on the $T-S$ curve as shown in Fig.\ref{fig2}, 
and $c_{\hat{Q}}$ can take negative values between the two singularities. 
This indicates that first order phase transitions can happen between the two 
singularities. At $\hat Q=\hat Q_c$, the two singularities merge together, 
the negative-valued branch of $c_{\hat{Q}}$ cease to appear, 
and the phase transition becomes critical. For $\hat Q>\hat Q_c$, the isochargge 
heat capacity no longer diverge at finite $T$, indicating that there is no
$T-S$ phase transitions in such cases. In the high temperature limit,
$c_{\hat{Q}}$ becomes independent of $\hat Q$, with asymptotic values given by
\begin{align}
\lim_{t\rightarrow \infty} c_{\hat{Q}} \sim 
\frac{384\sqrt{3} \pi}{125} \bfrac{\ell}{L}^3 t^3.
\label{cQd5}
\end{align}

The parameter $\alpha_{\hat{\Phi}}$ is naturally associated with isovoltage processes. 
We will be particularly interested in the cases with $\hat\Phi<\hat\Phi_{\rm thr}$, 
in which the minimum on the isovoltage $T-S$ curve exist and the chemical potential
versus temperature curve is branched. The analytical expression for $\alpha_{\hat{\Phi}}$
reads
\begin{align}
\alpha_{\hat{\Phi}}=\frac{20\sqrt{3}\pi\ell}{\phi^2+10s^{2/3}-15}
=\frac{4\tilde{s}^{1/3}}{T_{\rm min}(\tilde{s}^{2/3}-1)},
\end{align}
where the first expression is applicable to any admissible $\hat \Phi$, while the last
expression is applicable only for $\hat\Phi<\hat\Phi_{\rm thr}$, 
because otherwise $T_{\rm min}$ will be either zero or undefined.
The parameter $\beta_{\hat{Q}}$ is associated with isocharge processes. Therefore 
we adopt the relative parameters $s, t$ to evaluate $\beta_{\hat{Q}}$,
\begin{align}
\beta_{\hat{Q}} &=-\frac{4 \sqrt{3} \pi  \ell s^{5/3}}{q^2-3 s^{4/3}+2 s^2}.
\end{align}
As for the parameter $\kappa_T$, it is naturally associated with isothermal processes.
$\kappa_T$ can be written as an explicit function in $(s,t)$,
\begin{align}
\kappa_T&=-\frac{15 \sqrt{30} L^{3/2} 
\left(s^{2/3}-2 s^{1/3} t+1\right)}{s^{1/3} \left(5 s^{1/3}-6 t \right) 
\sqrt{10 s^{2/3}-24 s^{1/3} t+15}}.
\end{align}
However from the definition, it is better to think of $\kappa_T$ as a function in 
$(t,\phi)$. Therefore the variable $s$ in the above equation 
should be regarded as an implicit 
function in $(t,\phi)$ by use of eqs.\eqref{equationofstate} and \eqref{phiqeq}.

\begin{figure}[ht]
\begin{center}
\includegraphics[width=.35\textwidth]{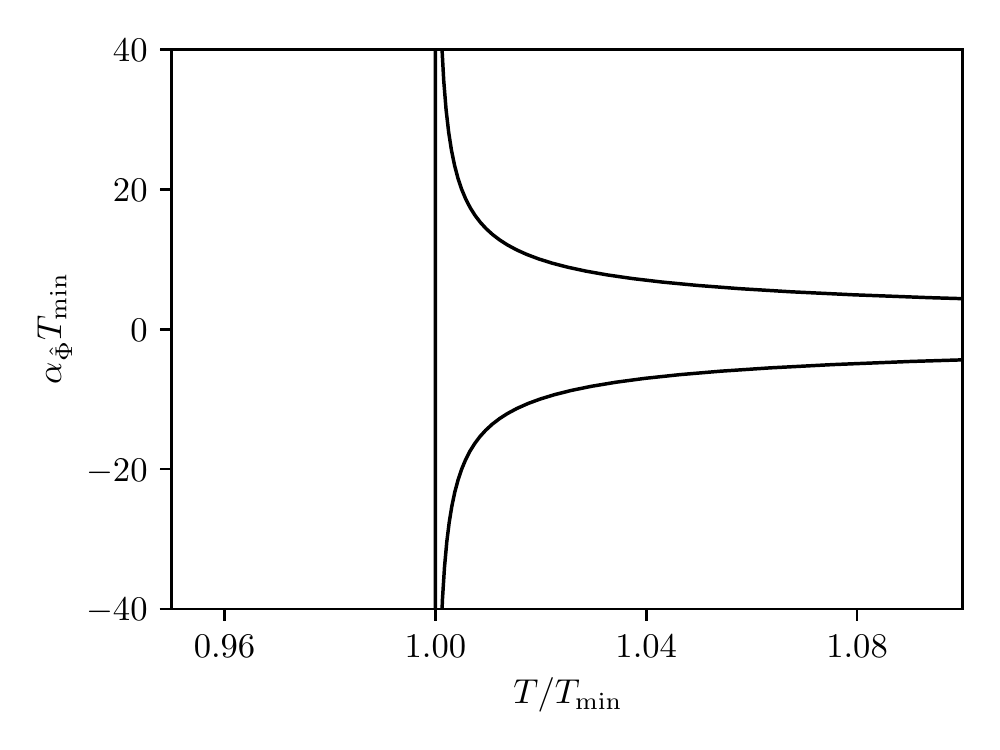}\hspace{.5em}
\includegraphics[width=.35\textwidth]{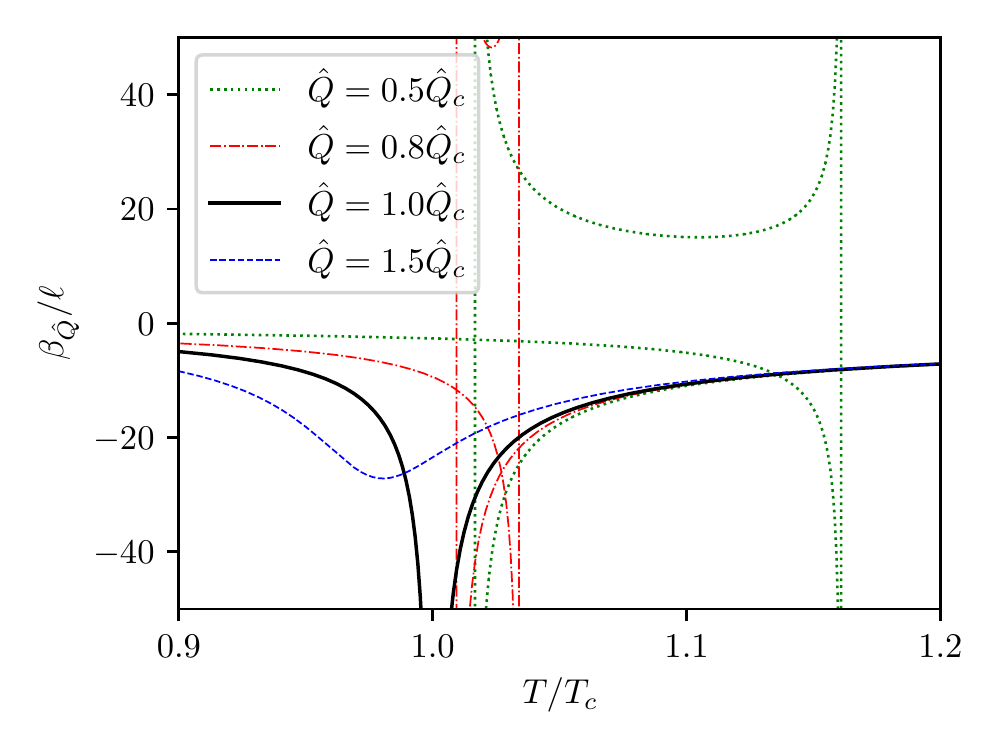}
\caption{The $\alpha_{\hat{\Phi}}-T$ and $\beta_{\hat{Q}}-T$ curves}
\label{fig6}
\end{center}
\end{figure}

The plots of the parameters $\alpha_{\hat{\Phi}}$ and $\beta_{\hat{Q}}$ versus $T$ 
are presented in Fig.\ref{fig6}. The singularity of $\alpha_{\hat\Phi}$ corresponds 
to the minimum of the isovoltage $T-S$ curve. The two singularities of 
$\beta_{\hat Q}$ at $\hat Q<\hat Q_c$ corresponds to the two extrema on the 
isocharge $T-S$ curve, which merge together at $\hat Q=\hat Q_c$. 

\begin{figure}[ht]
\begin{center}
\includegraphics[width=.35\textwidth]{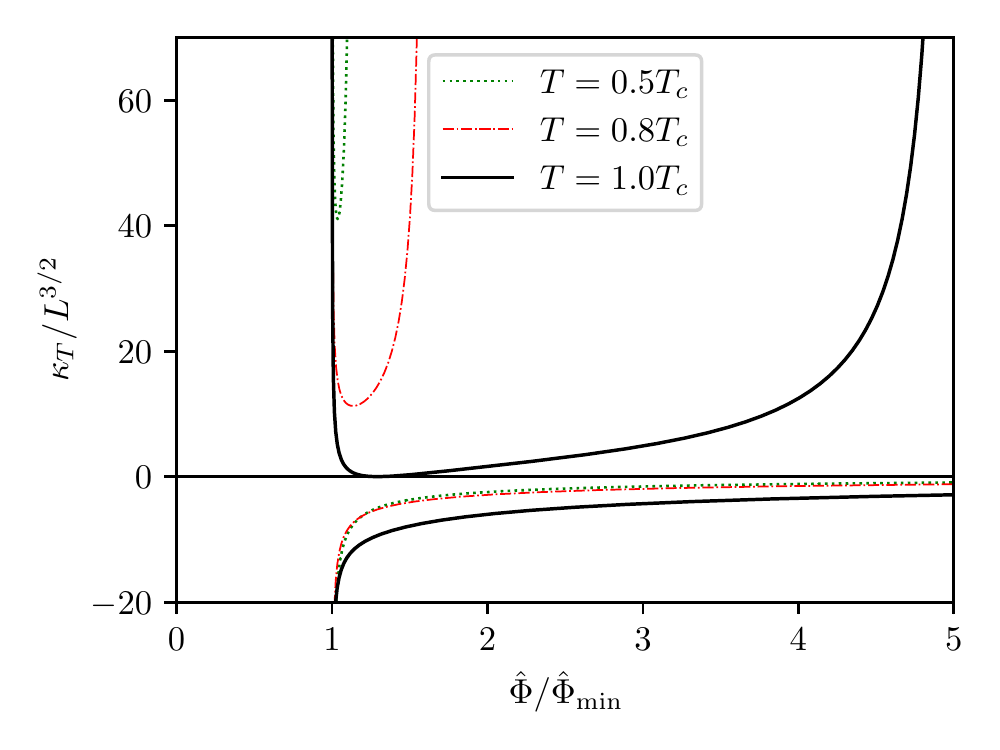}\hspace{1em}
\includegraphics[width=.35\textwidth]{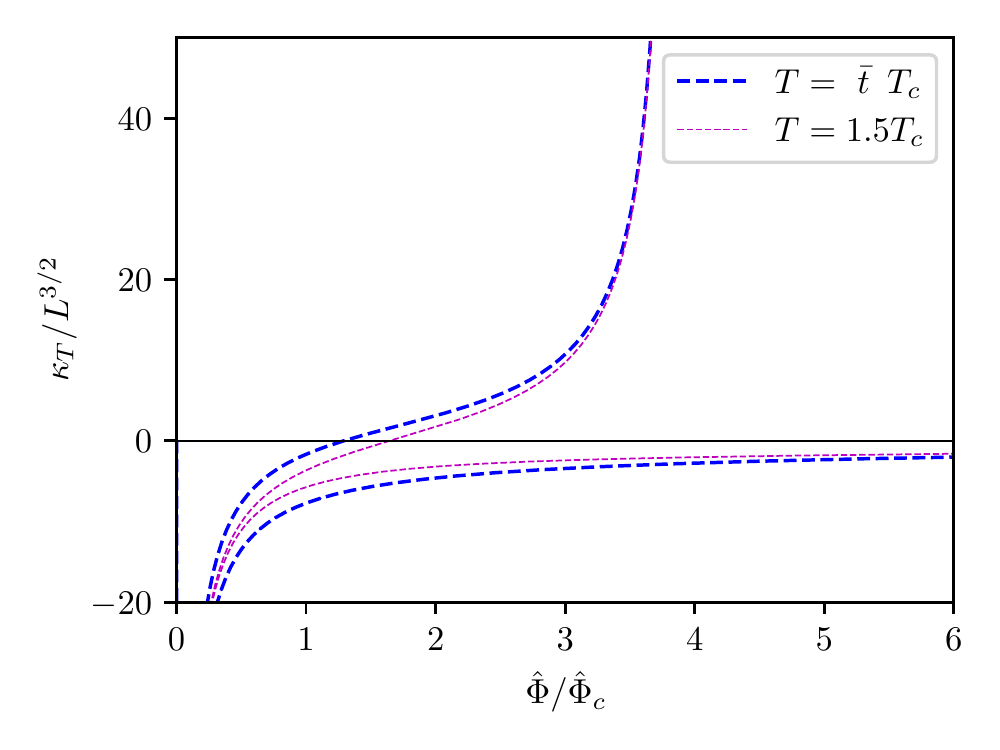}
\caption{ $\kappa_T-\hat{\Phi}$ curves for $t<\bar{t}$ and for $t \ge \bar{t}$}
\label{figka}
\end{center}
\end{figure}

The plots for the parameter $\kappa_T$ versus $\hat\Phi$ are more subtle, and we 
present the result separately in Fig.\ref{figka}. The left picture in 
Fig.\ref{figka} contains only the curves with $t<\bar t$, 
wherein we intentionally rescaled $\hat\Phi$ with $\hat\Phi_{\rm min}$ 
in order to indicate that the divergence of $\kappa_T$ in such cases 
corresponds to the minimum $\hat\Phi_{\rm min}$ appeared in the 
isovoltage $\hat\Phi-\hat Q$ curves. In such cases, the $\kappa_T$ 
values are branched, 
wherein the negative branch corresponds to the part of the part of the 
isothermal $\hat\Phi-\hat Q$ curve at $\hat Q< \hat Q_{\rm min}$, 
and the positive branch corresponds to the part of the part of the 
isothermal $\hat\Phi-\hat Q$ curve at $\hat Q>\hat Q_{\rm min}$. 
The right picture in 
Fig.\ref{figka} contains only the curves with $t\geq \bar t$, 
in which cases $\hat\Phi_{\rm min}$ is either zero or does not exist. 
Therefore the curves in these cases are created by scaling 
$\hat\Phi$ with $\hat\Phi_c$. It should be noticed that, for 
$t\geq \bar t$, $\kappa_T$ does not have any singularity at 
nonzero $\hat\Phi$. That the $\kappa_T-\hat\Phi$ curves are still
branched is due to the fact that the isothermal $\hat\Phi-\hat Q$ relation 
is already multivalued, as shown on the right plots in Fig.\ref{phi-q}.

\subsubsection{Scaling properties near criticality and critical exponents}

Scaling properties near the critical point is important in distinguishing different
universality classes in thermodynamics. In the present model, the only critical point
appears in the isocharge $T-S$ process. In order to compare with the $P-V$ 
criticalities appearing in systems consisted of ordinary matter, we need to make a 
list of parameter correspondences. The parameters $P, V, T$ for ordinary matter 
corresponds to $T, S, \hat Q$ in the present case in exactly the same order. 
This means that when considering critical exponents, the isochoric heat capacity $C_V$
for ordinary matter corresponds not to $C_{\hat Q}$ but rather to 
$\kappa_S^{-1} \equiv -{\hat{Q}}\pfrac{\hat{\Phi}}{{\hat{Q}}}_S$,
where $\kappa_S$ is similar to $\kappa_T$ defined in eq.\eqref{abc}, 
but is associated with adiabatic processes. Similarly, 
we expect that the scaling behavior of the isothermal compression 
coefficient $\kappa_T$ ({\it not the one defined in eq.\eqref{abc}}) 
for ordinary matter corresponds to that of $C_{\hat Q}$ in our 
case, etc. Below we shall present the detailed studies.

Writing 
\begin{align}
s=S/S_{c} \equiv 1+\delta s, \quad t=T/T_{c} \equiv 1+\delta t, \quad 
q={\hat{Q}}/{\hat{Q}}_{c} \equiv 1+\delta q,
\end{align}
we get from eq.\eqref{equationofstate} the following relation,
\begin{align}
\delta t=\frac{5}{162}(\delta s)^3, 
\end{align}
which means 
\begin{align}
|T-T_c|\sim |S-S_c|^\delta,\quad \delta=3.
\end{align}
Near the critical point {\cite{tong2012lectures}}, we have 
\begin{align}
t(q,1+\delta s) = t(q,1-\delta s),
\end{align}
which is implied by the fact that the critical phase transition is approached 
via an isocharge isothermal process. The above identity implies
\begin{align}
\delta q=-\frac{2}{9}(\delta s)^2,
\end{align}
i.e.
\begin{align}
|S_{\rm large}-S_{\rm small}| \sim \left|{\hat{Q}}-{\hat{Q}}_c\right|^{\beta},
\quad
\beta=\frac{1}{2}.
\end{align}

Using the definition of the isocharge heat capacity, we get
\begin{align}
C_{\hat{Q}} &= T\pfrac{S}{T}_{\hat{Q}} 
\propto t \pfrac{\delta s}{\delta t}_{\delta q}
\propto (\delta t)^{-{2}/{3}}
\propto (\delta s)^{-2}
\propto -(\delta q)^{-1}, 
\end{align}
When the charge approaches the critical value from below (i.e. $\delta q\to 0^-$), 
we have $C_{\hat{Q}} \rightarrow \infty$. 
Thus we get the third critical exponent $\gamma=1$.

Now we turn to look at the behavior of $\kappa_S^{-1} = -{\hat{Q}}
\pfrac{\hat{\Phi}}{{\hat{Q}}}_S$. Since $\hat{\Phi}=\pfrac{M}{{\hat{Q}}}_S$, we have
\begin{align}
\kappa_S^{-1}=-{\hat{Q}}\pfn{M}{{\hat{Q}}}{2}_S
=-\frac{M_c}{{\hat{Q}}_c}q\pfn{m}{q}{2}_s.
\label{ksn1}
\end{align}
When $q\rightarrow 1^+$, we can make use of the caloric equation of states \eqref{ceos} 
to get
\begin{align}
\kappa_{S+}^{-1} = -\frac{1}{\sqrt{30 } L^{3/2} } \propto (\delta q)^0.
\label{ksn1p}
\end{align}
When $q\rightarrow 1^-$, the caloric equation of states \eqref{ceos} is no longer reliable, 
so we adopt Maxwell's linear interpolation {\cite{thompson1988classical}}
\begin{align}
\bar{m}(s,q)=m(s_{\rm small},q)+(s-s_{\rm small})
\frac{m(s_{\rm large},q)-m(s_{\rm small},q)}{s_{\rm large}-s_{\rm small}},
\label{mbar}
\end{align}
where $m(s,q)$ on the right hand side is given by the original caloric equation 
of state \eqref{ceos}. When approaching the critical point, we have
\[
\epsilon \equiv \frac{1}{2}(s_{\rm large}-s_{\rm small})= \delta s = 
\frac{3\sqrt{2}}{2}(-\delta q)^{1/2}=\frac{3\sqrt{2}}{2}(1-q)^{1/2}.
\]
Thus we can expand the right hand side of eq.\eqref{mbar} into power series in $\epsilon$
near the critical point and get
\[
\bar{m}(s,q)|_{s=1} =m(s,q)|_{s=1} +\frac{\epsilon^2}{2!}\pfn{m}{s}{2}_{s=q=1}
+\frac{\epsilon^4}{4!}\pfn{m}{s}{4}_{s=q=1}+\cdots.
\]
Notice that the $q$ dependence of the second and the third terms are encoded 
in $\epsilon$. Finally, using eq.\eqref{ksn1}, we get 
\begin{align}
\kappa_{S-}^{-1} &= -\frac{M_c}{{\hat{Q}}_c}q\pfn{\bar{m}}{q}{2}_{s=1}
=-\frac{M_c}{{\hat{Q}}_c}q\pfn{m}{q}{2}_{s=1}
-\frac{27M_c}{16{\hat{Q}}_c} q \pfn{m}{s}{4}_{s=q=1}\nonumber\\
&=-\frac{1}{\sqrt{30} L^{3/2} }-\frac{\sqrt{30}}{12 L^{{3/2}}} \propto (\delta q)^0.
\label{ksn1n}
\end{align}
Combining eqs.\eqref{ksn1p} and \eqref{ksn1n} we find that
\[
\kappa^{-1}_{S\pm} \propto (\delta q)^{-\alpha_\pm},\quad \alpha_\pm=0,
\]
and there is a finite jump between $\kappa^{-1}_{S\pm}$:
\[
\kappa^{-1}_{S+}-\kappa^{-1}_{S-}= \frac{\sqrt{30}}{12 L^{{3/2}}}.
\]
This jump arises because of the existence of a discontinuity in the second 
order derivatives of $M$ with respect
to $\hat Q$, i.e. $\pfn{M}{{\hat{Q}}}{2} \big|_{{\hat{Q}}\rightarrow {\hat{Q}}_c^-} 
\neq \pfn{M}{{\hat{Q}}}{2} \big|_{{\hat{Q}}\rightarrow {\hat{Q}}_c^+}$, 
which indicates that the critical point corresponds to a second order phase transition
in terms oh Ehrenfest classification.

It may be illustrative to summarize the above results about the critical exponents
in conjunction with a comparison with the corresponding scaling relations for 
ordinary matter. Explicitly, we have found a perfect correspondence 
between the scaling relations for ordinary matter and for the charged AdS black hole,
\[
\left\{\begin{array}{l}
	C_V \propto {\begin{array}{l}
			|T-T_c|^{-\alpha^-} \\
			|T-T_c|^{-\alpha^+}
	\end{array}},\\
	\rho_l-\rho_g \propto |T-T_c|^{\beta},\\
	\kappa_T \propto |T-T_c|^{-\gamma},\\
	|P-P_c|\propto |\rho-\rho_c|^\delta,
\end{array}\right.
\Longleftrightarrow
\left\{\begin{array}{l}
	\kappa_S^{-1} \propto {\begin{array}{l}
			|{\hat{Q}}-{\hat{Q}}_c|^{-\alpha^-} \\
			|{\hat{Q}}-{\hat{Q}}_c|^{-\alpha^+}
	\end{array}},\\
	S_{\rm large}-S_{\rm small} \propto |T-T_c|^{\beta},\\
	C_{\hat{Q}} \propto |{\hat{Q}}-{\hat{Q}}_c|^{-\gamma},\\
	|T-T_c|\propto |S_{\rm large}-S_{\rm small}|^\delta,
\end{array}\right.
\]
where $\alpha_\pm=0$, $\beta=\frac{1}{2}$, $\gamma=1$ and $\delta =3$. Naturally, 
they obey the same scaling symmetry as in ordinary thermodynamic systems {\cite{pathria2016statistical}}, 
$$
2-\alpha =2\beta +\gamma =\beta(\delta+1 ) =\gamma (\delta +1)/(\delta -1)
$$

\subsection{$(5,2)$-model: CS gravity coupled to Maxwell field}

The $(5,2)$-model is a CS like model for which the total action takes the form
\begin{align}
\mathcal{A}_{(5,2)}&=\frac{1}{6 \pi^{2} \ell^2 G_{(5,2)}}  
\int\left(R+\frac{6}{\ell^{2}}+\frac{\ell^2}{4}\mathcal{R}^2 \right) \sqrt{-g} \rd^5 x 
-\frac{1}{8\pi^2} \int F_{\mu\nu}F^{\mu\nu} \sqrt{-g} \rd^5 x,
\end{align}
where
\[
\mathcal{R}^2= R_{\mu \nu \gamma \delta} R^{\mu \nu \gamma \delta}
-4 R_{\mu \nu} R^{\mu \nu}+R^{2}.
\]
The variation of the total action reads
\[
\delta \mathcal{A}_{(5,2)}=E^{g}_{\mu\nu} \delta g^{\mu\nu}
+E^\mu_A \delta A_\mu + \rd\boldsymbol{\Theta},
\]
where
\begin{align}
&E^{g}_{\mu\nu} =R_{\mu \nu}-\frac{1}{2} g_{\mu \nu} R-\frac{3}{\ell^{2}} g_{\mu \nu} 
+\frac{\ell^2}{4}H_{\mu\nu} -3 \pi^{2} \ell^2 G_{(5,2)} T_{\mu \nu}, \\
&E^\mu_A = \nabla_\nu F^{\mu\nu},
\end{align}
with
\[
T_{\mu\nu} = \frac{1}{2\pi^2}(F_{\mu\rho}F_{\nu}^{\rho}
-\frac{1}{4}g_{\mu\nu}F_{\rho\sigma}F^{\rho\sigma}),
\]
\[
H_{\mu \nu} =2\left(R_{\mu \lambda \rho \sigma} R_{\nu}{}^{\lambda \rho \sigma}
-2 R_{\mu \rho \nu \sigma} R^{\rho \sigma}-2 R_{\mu \sigma} R_{\nu}^{\sigma}
+R R_{\mu \nu}\right) 
-\frac{1}{2}\left(R_{\mu \nu \gamma \delta} R^{\mu \nu \gamma \delta}
-4 R_{\mu \nu} R^{\mu \nu}+R^{2}\right) g_{\mu \nu}.
\]
$\boldsymbol{\Theta}$ is a $\delta g^{\mu\nu}$-dependent four form 
field which encodes the on-shell Noether current if $\delta g^{\mu\nu}$ is replaced by the
transformation of $g^{\mu\nu}$ under diffeomorphism symmetries.
The associated Noether charge can be used to evaluate the Wald entropy,
which reads
\[
S=\frac{4 \pi  r_h \left(3 \sqrt{1+2 G_{(5,2)} M-\frac{G_{(5,2)} Q^2}{\pi  r_h^2}}
-\frac{2 r_h^2}{\ell^2}\right)}{3 G_{(5,2)}}.
\]

The $\hat Q,\hat\Phi$ and $\mu, N$ variables can be inferred from the 
generic definitions given in Sec.2 by inserting $d=5, k=2$ and the thermal and 
caloric equations of states for this particular case are worked out to be
\begin{align}
T&=\dfrac{\pi \mathcal{Z} \left[\dfrac{\left(\left(\ell^2 \mathcal{Z} \right)^{2/3}
-4 \pi ^{2/3} \ell^2\right)^4}{4 \pi ^{4/3} \left(\ell^2 \mathcal{Z} \right)^{4/3}}
-\ell^2 L \mathcal{Q}^2+\dfrac{\left(\left(\ell^2 \mathcal{Z} \right)^{2/3}
-4 \pi ^{2/3} \ell^2\right)^6}{16 \pi ^2 \ell^6 \mathcal{Z} ^2}\right]}
{\left[\left(\ell \mathcal{Z} \right)^{2/3}-4 \pi ^{2/3} \ell^{4/3}\right]^3 
\left[\frac{1}{4} \sqrt[3]{\pi } \left(\ell^2 \mathcal{Z} \right)^{2/3}
+\pi  \ell^2 \left(4 \pi ^{2/3} \mathcal{Z}^{-2/3}-1\right)\right]},
\label{tsd5k2}
\\
M&=N\dfrac{(\pi\mathcal{Z})^{2/3} 
\left[\dfrac{\left(\left( \ell^2 \mathcal{Z}\right)^{2/3}
-4 \pi ^{2/3} \ell^2\right)^6}{\pi ^2  \ell^4 \mathcal{Z}^2}
+\dfrac{8 \left(\left( \ell^2 \mathcal{Z}\right)^{2/3}
-4 \pi ^{2/3} \ell^2\right)^4}{\pi ^{4/3} \ell^{2/3} \mathcal{Z}^{4/3}}
+32 \ell^4 L \mathcal{Q}^2\right]}{32 \ell^{16/3} L \left(\mathcal{Z}^{2/3}
-4 \pi ^{2/3} \ell^{2/3}\right)^2},
\end{align}
where 
\[
\mathcal{Z}=\sqrt{64\pi^2 \ell^2+9L^2 \mathcal{S}^2}+3L\mathcal{S},\quad
\mathcal{S}=S/N,\quad \mathcal{Q}=Q/N.
\]
These equations of states indicate that $T$ and $M$ are respectively zeroth and 
first order homogeneous functions in the extensive variables. 
Moreover, non-negativeness of the temperature gives rise to an upper bound for 
$\hat Q$, which reads
\[
\hat{Q}\leq \frac{ \left[ (\ell^2 N^2\mathcal{Z})^{2/3}
-4\pi^{2/3} \ell^2 N^2 \right] 
\left[16\pi^{4/3}\ell^2 N^2  -4\pi^{2/3} (\ell N)^{4/3}\mathcal{Z}^{2/3} 
+ (\ell N)^{2/3}\mathcal{Z}^{4/3}\right]^{1/2}}{4\pi \sqrt{L}\ell^3 N^3 \mathcal{Z}},
\]
which is to be understood as the Bogomol'nyi bound for the present case.

The overwhelming complicated form of eq.\eqref{tsd5k2} 
prevents us from analyzing the thermodynamic processes right away from
the thermal equation of states. To proceed, we choose to rewrite the 
thermal equation of states $T=T(S,\hat Q, N)$ in a parametric from
\begin{align}
T(r_h,\hat{Q},N)&=\frac{-\frac{\ell^2 L \hat{Q}^2}{N^2}
+\frac{4 r_h^6}{\ell^2}+4 r_h^4}{8 \pi  r_h^3 \left(\ell^2+r_h^2\right)},
\label{paramTQ}\\
S(r_h,N)&=\frac{4 \pi}{L}  N r_h \left(1+\frac{r_h^2}{3 \ell^2}\right),
\label{paramSQ}
\end{align}
where the radius $r_h$ of the black hole event horizon is taken as an intermediate 
parameter, while the dependence on the thermodynamic parameters $\hat Q$ and $N$ 
are made explicit by employing the simple relations
\[
G_{(5,2)} =\frac{L}{N} \qquad Q=\frac{\hat{Q}}{\sqrt{N}}.
\]
Notice that both $T(r_h,\hat{Q},N)$ and $S(r_h,\hat{Q},N)$ are single-valued in $r_h$
at constant $\hat Q$ and $N$.

The above parametric form of the thermal equation of states allows us to 
analyze the isocharge processes of the model. By straightforward calculations, 
we can get
\begin{align}
\pfrac{T}{S}_{\hat{Q}} =\frac{\displaystyle \pfrac{T}{r_h}_{\hat{Q},N}}
{\displaystyle \pfrac{S}{r_h}_{\hat{Q},N}}
=\frac{L \left[\ell^4 \left(4 N^2 r_h^4+5 L \hat{Q}^2 r_h^2\right)
+8 N^2 \ell^2 r_h^6+4 N^2 r_h^8+3 \ell^6 L \hat{Q}^2\right]}
{32 \pi^2 N^3 r_h^4 \left(\ell^2+r_h^2\right)^3}.
\end{align}
Though the expression still looks complicated, it can be easily seen that 
$\displaystyle \pfrac{T}{S}_{\hat{Q}}$ is identically positive. This means that 
$\displaystyle C_{\hat{Q}}=T\pfrac{S}{T}_{\hat{Q}}>0$, which means that there is only a single 
stable AdS black hole phase in the isocharge processes for the present model, no 
$T-S$ phase transition could occur. 
\blue{It can be shown that at high temperatures, the asymptotic value of 
$c_{\hat{Q}}=C_{\hat{Q}}/N$ for this model is}
\begin{equation}
\blue{\lim_{t\rightarrow \infty} c_{\hat{Q}} =\frac{32 \pi ^4 \ell^4 }{L} T^3}.
\end{equation}

In order to describe the isovoltage processes, we need to employ the relation
$\hat Q = 2 N\hat{\Phi} r_h^2$ to rearrange eq.\eqref{paramTQ} into 
the form
\begin{align}
T(r_h,\hat{\Phi})&= r_h\frac{\ell^2 +r_h^2 
-\ell^4 L \hat{\Phi}^2}{2 \pi \ell^2  (\ell^2+r_h^2)}.
\label{paramTP}
\end{align}
It is then easy to get
\[
\pfrac{T}{S}_{\hat{\Phi}} 
=\frac{\displaystyle \pfrac{T}{r_h}_{\hat{\Phi},N}}
{\displaystyle \pfrac{S}{r_h}_{\hat{\Phi},N}}
= \frac{L \left[-\ell^6 L \hat{\Phi}^2+\ell^4 \left(L \hat{\Phi}^2 r_h^2+1\right)
+2 \ell^2 r_h^2+r_h^4\right]}{8 \pi ^2 \hat{\Phi} \left(\ell^2+r_h^2\right)^3}.
\]
The equation $\displaystyle \pfrac{T}{S}_{\hat{\Phi}}=0$ 
with the condition $\displaystyle \pfn{T}{S}{2}_{\hat{\Phi}}>0$ 
has a single solution
\[
r_h=r_{\rm min}=\frac{\ell}{\sqrt{2}}
\left[ \ell \hat{\Phi} \left(\ell^2 L^2 \hat{\Phi}^2+8L\right)^{1/2} 
-\ell^2 L \hat{\Phi}^2-2 \right]^{1/2},
\]
however, the real-valuedness of $r_{\rm min}$ requires 
$\hat{\Phi}\geq \frac{1}{\ell \sqrt{L}}$, which in turn implies that 
the minimum at $r_h=r_{\rm min}$ corresponds to a non-positive temperature and 
is physically irrelevant. For all states with $T>0$, it is not difficult to check that 
$\displaystyle \pfrac{T}{S}_{\hat{\Phi}}$ is always positive, therefore there is 
no isovoltage $T-S$ transitions in the present model.

We can also write down the parametrized $\mu-T$ equation of states, which read
\begin{align}
\mu(r_h,\hat{\Phi})&=-\frac{r_h^2 \left[\ell^4 \left(4 L \hat{\Phi}^2 r_h^2+6 \pi \right)
+7 \pi  \ell^2 r_h^2+\pi  r_h^4 -12 \ell^6 L \hat{\Phi}^2\right]}
{6 \pi  \ell^7 \left(\ell^2+r_h^2\right)},
\label{paramu}
\end{align}
where implicit dependence on $T$ is implied by use of eq.\eqref{paramTP}. 
Then it is straightforward to get
\begin{align}
\pfrac{\mu}{T}_{\hat{\Phi}} =\frac{\displaystyle \pfrac{\mu}{r_h}_{\hat{\Phi}}}
{\displaystyle\pfrac{T}{r_h}_{\hat{\Phi}}}
=-\frac{4 \pi  \left(3 \ell^2 r_h +r_h ^3\right)}{3 \ell^5},
\end{align}
which is identically negative at all $r_h>0$, which means that $\mu$ is 
monotonically decreasing with $T$. From eq.\eqref{paramTP} one can see that at any
fixed $T$, $\hat\Phi$ can be regarded as an implicit function in $r_h$. Thus one 
can solve eq.\eqref{paramTP} at $T=0$ as an equation for $\hat\Phi$ and 
substitute the result into eq.\eqref{paramu}
to get
\[
\mu|_{T=0} = -\frac{r_h^2 \left[6(\pi-2)\ell^2+(4+\pi)r_h^2\right]}{6\pi\ell^7}<0.
\]
This result combined with the fact that $\displaystyle \pfrac{\mu}{T}_{\hat{\Phi}}<0$ 
indicate that $\mu$ is always negative and thus has no zeros at positive $T$. 
This excludes the possibility for the appearance of Hawking-Page like transitions 
in the present model. 

Summarizing this subsection, we have seen that the thermodynamic behavior of the 
$(5,2)$-model is in sharp contrast to that of the $(5,1)$-model. No phase 
transition of any type (be it isocharge, isovoltage or Hawking-Page like) 
could occur in the $(5,2)$-model. We believe this is also the case in 
any CS like or $(2k+1,k)$-models.

\subsection{$(6,2)$-model: BI gravity coupled to Maxwell field}

The full action for the $(6,2)$-model reads
\begin{align}
\mathcal{A}_{(6,2)}=\frac{3}{32 \pi^{2} \ell^2 G_{(6,2)}}  
\int\left(R+\frac{10}{\ell^{2}}+\frac{\ell^2}{12}\mathcal{R}^2 \right) \sqrt{-g} \rd^6 x 
-\frac{3}{32\pi^2} \int F_{\mu\nu}F^{\mu\nu} \sqrt{-g} \rd^6 x.
\end{align}
Except for the different choices of coupling coefficients and the different 
spacetime dimensions, the above action is quite similar to that of the $(5,2)$-model. 

The Wald entropy for the charged spherically symmetric AdS black hole solution 
in this case reads
\[
S=\dfrac{\pi  \left(\ell^2 \sqrt{G_{(6,2)} \left(8 M r_h^3-\dfrac{2 Q^2}{\pi }\right)}
-r_h^4\right)}{G_{(6,2)} \ell^2},
\]
and the other thermodynamic parameters/functions can be inferred from the 
general definition presented in Sec.2. 

The thermal and caloric equations of states for this case can also be 
worked out analytically, which appears to be much simpler than the case
of $(5,2)$-model,
\begin{align}
T&=\frac{5 \mathcal{S}^{2} -14\pi^{1/2}\ell \mathcal{S}^{3 / 2}
+13\pi \ell^2 \mathcal{S} -4 \pi^{3 / 2}\ell^{3}\mathcal{S}^{1/2}
- \pi^{2} L^{2}\mathcal{Q}^{2}}{8\pi^{5 / 4} \ell^{3 / 2}\mathcal{S}^{1/2}
(\mathcal{S}^{1/2}-\pi^{1/2}\ell)^{5 / 2}},\\
M&=N \frac{\left(3 \mathcal{S}^{2} -6\pi^{1/2} \ell  \mathcal{S}^{3 / 2}
+3 \pi \ell^{2} \mathcal{S}+\pi^{2} L^{2} \mathcal{Q}^{2} \right)}
{6\pi^{5 / 4} \ell^{3 / 2} L^{2} (\mathcal{S}^{1/2}-\pi^{1/2}\ell)^{3 / 2}},
\end{align}
where $\mathcal{S}=L^2\frac{S }{N} +\pi\ell^2$ and $\mathcal{Q}=\frac{\hat{Q}}{N}$. 
The isocharge $T-S$ curves may contain an inflection point for $\hat Q$ below some
critical value $\hat Q_c$, and the values of various thermodynamic parameters/functions 
can be worked out analytically. However, the analytical expressions for the 
critical parameters are too complicated which do not deserve to be displayed here. 
As an alternative, we present their approximate numerical values below,
\begin{align}
T_c &\approx \frac{0.1747}{\ell}, \quad\quad ~
S_c\approx \frac{0.9058  \ell^2}{L^2}N,
\\
\hat{\Phi}_c &\approx \frac{0.2406}{\ell L},\quad \quad\,
\displaystyle \hat{Q}_c\approx \frac{0.0358 \ell^2}{L}N,
\\
M_c &\approx \frac{0.2410  \ell}{L^2}N, \quad
F_c \approx \frac{0.0828  \ell}{L^2}N. 
\end{align}
Critical values of quantities which are intensive are simply constants and 
those for extensive quantities are explicitly proportional to $N$, thus the 
extensivity and Euler homogeneity are transparent. 

The forthcoming analysis is in complete analogy to the case of $(5,1)$-model, 
and we shall only present some of the plots for brevity. 

\begin{figure}[ht]
\begin{center}
\includegraphics[width=.4\textwidth]{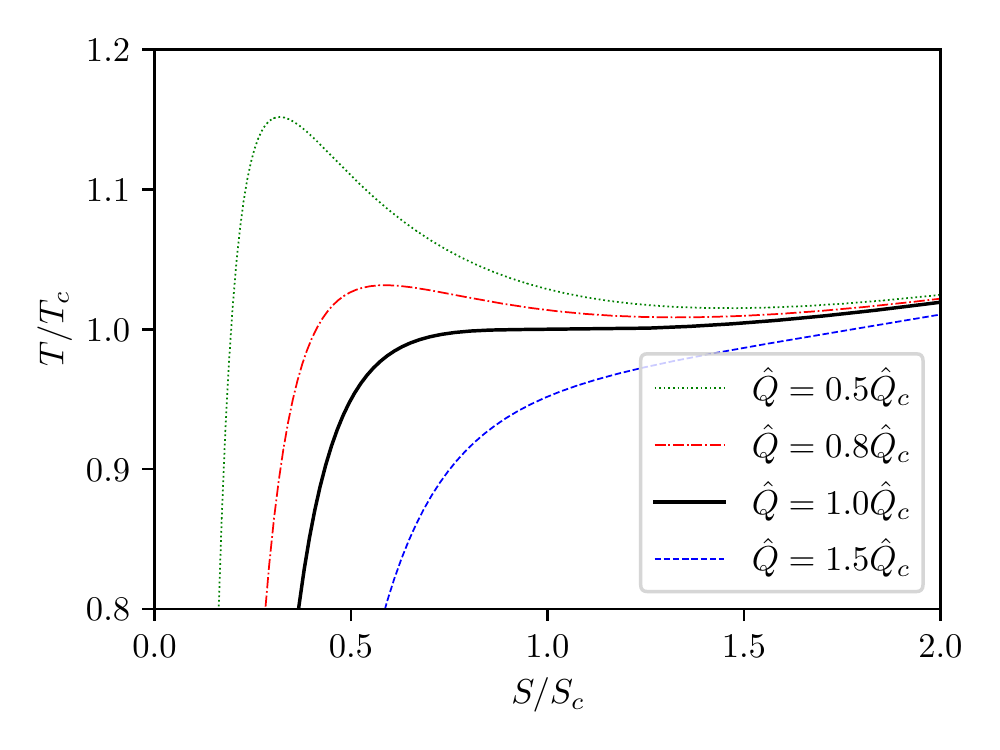}\hspace{1em}
\includegraphics[width=.4\textwidth]{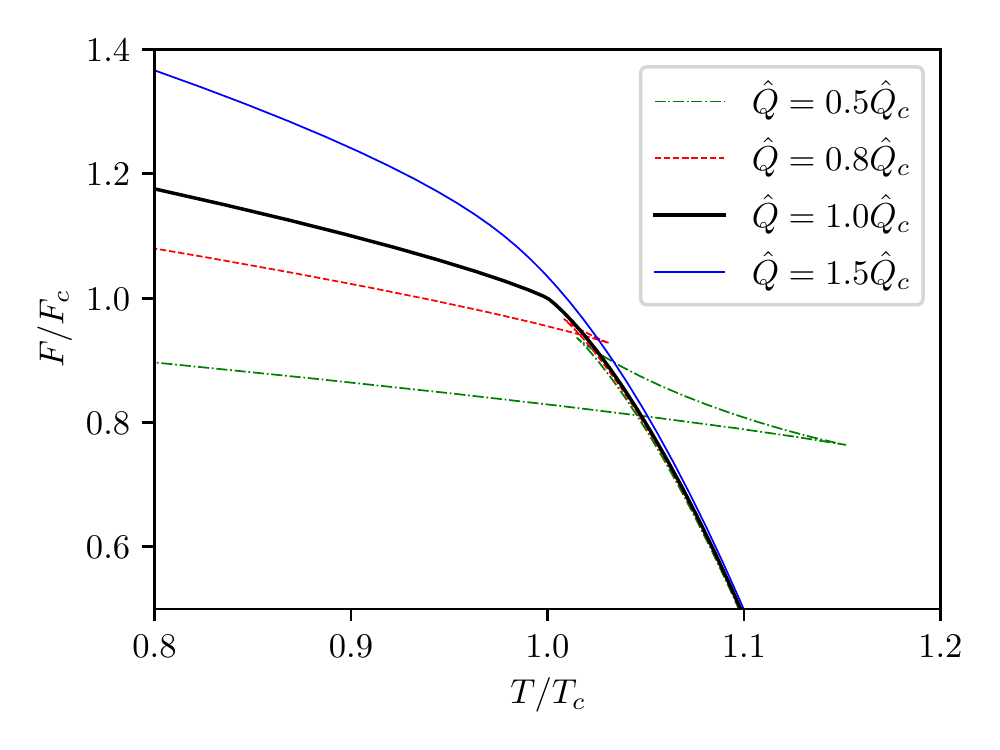}
\caption{Isocharge $T-S$ and $F-T$ curves for the $(6,2)$-model}\label{fig7}
\end{center}
\end{figure}

\begin{figure}[ht]
\begin{center}
\includegraphics[width=.4\textwidth]{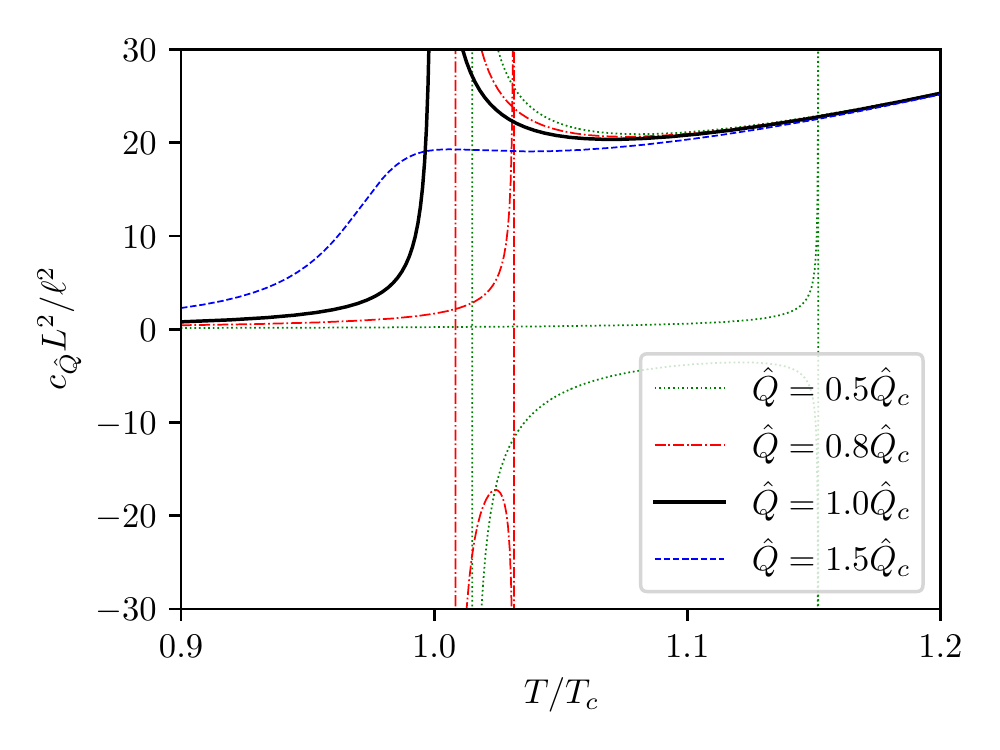}
\caption{Isocharge $c_{\hat Q}-S$ and $c_{\hat Q}-T$ curves 
for the $(6,2)$-model}\label{fig8}
\end{center}
\end{figure}

The isocharge $T-S$ and $F-T$ curves are presented in Fig.\ref{fig7},
and the isocharge $c_{\hat Q}-T$ curves are presented in 
Fig.\ref{fig8}. One can see that the curves presented in Fig.\ref{fig7} 
are very similar to the curves shown in Fig.\ref{fig2}
for the $(5,1)$-model, and the curves in Fig.\ref{fig8} are very similar 
to those presented in Fig.\ref{fig5}. It is not surprising that the 
isocharge $T-S$ behavior for the $(6,2)$-model is qualitatively identical 
to that of the $(5,1)$-model. 

At high temperatures, all $c_{\hat Q}-T$ curves in Fig.\ref{fig8} 
merge together, with the asymptotic value
\begin{align}
\lim_{t\rightarrow \infty} c_{\hat{Q}} \approx \frac{7.4670 \ell^2}{L^2} t^4,
\label{cQd6}
\end{align}
where again $t=T/T_c$.

By use of numeric techniques, it is not difficult to see that the behavior of 
the isovoltage $T-S$ processes in the $(6,2)$-model is also 
qualitatively similar to the case of $(5,1)$-model and hence we omit the 
corresponding plots. 

We also studied the behaviors of the process 
parameters $\alpha_{\hat{\Phi}}, \beta_{\hat{Q}}, \kappa_T$ and found that their 
qualitative behaviors are similar to the case of $(5,1)$-model. However, 
due to the more complicated form of the equations of states, the analytical 
expressions for these parameters are more involved and does not deserve to 
be displayed here. Even though, we found that the scaling properties of the 
$(6,2)$-model coincide exactly with the case of $(5,1)$-model, in the sense that the 
critical exponents are exactly the same. Therefore, we conclude that the 
$(6,2)$-model and the $(5,1)$-model belong to the same universality class, 
although they belong to different types of gravity models 
(BI and EH types respectively).

\section{Summary and discussions} 

We have shown that the RPST formalism is applicable to the large class of 
$(d,k)$-models and yields the correct Euler and Gibbs-Duhem relations which guarantee 
the extensivity of the relevant thermodynamic systems. 
\red{The applicability of the RPST formalism to the $(d,k)$-models makes us 
expect that the formalism may be universally applicable to all 
models of gravity. One reason to support this expectation lies in that, unlike the EPST formalism which extends the space of macroscopic states for black holes by the inclusion of cosmological constant or some other coupling coefficients which may be either present or absent in different models of gravity, the RPST formalism uses the Newton constant as an extra dimension in the space of macroscopic states for black holes, and the Newton constant must be present in all models of gravity and can be made as an overall factor of the total action by some simple field rescaling (e.g. the rescaling made in eq.\eqref{qhat} can be understood as a rescaling of the Maxwell field $A_{\mu}$.)}

By choosing the $(5,1)$-, $(5,2)$- 
and the $(6,2)$-models as representatives of the classes of EH like, CS like and BI like 
gravity models coupled with Maxwell field, we found that the $(5,1)$- and the 
$(6,2)$-models have basically the same thermodynamic behaviors, e.g. the
appearance of the supercritical isocharge $T-S$ phase transitions at temperatures above 
a critical temperature, the existence of the isovoltage $T-S$ transitions above some minimum
temperature which do not subject to any equilibrium conditions, 
the occurrence  of Hawking-Page like transitions and the exactly same scaling 
properties at the critical point, etc. These behaviors were also observed for 
four dimensional RN black hole and Kerr-AdS black holes in our previous 
works \red{\cite{gao2021restricted,gao2022thermodynamics}}, 
and it seems that the EH and BI like models belong to the same universality
class under the RPST formalism. The $(5,2)$-model, however, behaves very differently. 
There is no phase transitions of any type in this model, which indicates that the CS 
like models may belong to a different universality class under the RPST formalism\footnote{\red{The result of \cite{wb}, which studies the original Chern-Simons gravity, i.e. Einstein-Hilbert gravity in (2+1)-dimensions, also supports this conclusion.}}. 
These results are a first known application of the RPST formalism to the 
cases of higher curvature gravity models. 
We expect that further applications of the RPST formalism 
to other higher curvature gravity models may help to classify the possible universality classes
in all gravity models.

\red{Regarding the possible phase transitions in the cases of EH and BI 
like models, let us stress that, although there are similar 
near critical scaling properties to the cases of Van der Waals 
liquid-gas phase transitions, the $T-S$ phase transitions do not 
involve any changes in volume or pressure. The occurrence of the $T-S$ 
transition is a simple consequence of the fact that some of the 
thermodynamic parameters (e.g. the temperature) of the black holes are not 
monotonic functions in the geometric parameters (e.g. the radius of the 
event horizon). Therefore, if we characterize the black hole states 
using the thermodynamic parameters, black holes of different geometric sizes
may appear to be in the same thermodynamic states, and that is why 
black hole phase transitions could occur.}

Among the various concrete behaviors of the example cases, 
the high temperature asymptotic behavior of the isocharge heat capacity
deserves a special mention. In all three example cases,
the high temperature asymptotic values of $c_{\hat{Q}}$ behave as
\[
c_{\hat{Q}}\sim T^{d-2}.
\]
Remember that $d-2$ can be regarded as the ``spatial dimension'' of the 
black hole event horizon regarded as a thermodynamic system. 
The high temperature asymptotic behavior $c_{\hat{Q}}\sim T^{d-2}$ for black holes
reminds us of the low temperature asymptotic 
behavior of the isochoric 
specific heat capacity $c_V$ for ordinary non-metallic solid matter, 
which is, according to Debye's theory, $c_V\sim T^D$, with $D$ being the spatial
dimension of the system. This similarity may not be a coincidence, there might be
some deep physics in behind.

\section*{Acknowledgement}
This work is supported by the National Natural Science Foundation of 
China under the grant No. 11575088.

\providecommand{\href}[2]{#2}\begingroup%\raggedright
\footnotesize\itemsep=0pt
\providecommand{\eprint}[2][]{\href{http://arxiv.org/abs/#2}{arXiv:#2}}

\end{document}